\documentclass[12pt]{article}

\usepackage{amstext}
\usepackage{latexsym}
\usepackage{amssymb}
\usepackage{tensor} 
\usepackage{amsmath}
\usepackage{graphicx}
\usepackage{ifpdf}
\usepackage{color}
\usepackage{float}

\def\R{\mathbb R}
\def\C{\mathbb C}
\def\N{\mathbb N}
\def\Z{\mathbb Z}

\newcommand{\eps}{\varepsilon}

\begin{document}

\newtheorem{theorem}{Theorem}[section]
\renewcommand{\thetheorem}{\arabic{section}.\arabic{theorem}}
\newtheorem{definition}[theorem]{Definition}
\newtheorem{deflem}[theorem]{Definition and Lemma}
\newtheorem{lemma}[theorem]{Lemma}
\newtheorem{example}[theorem]{Example}
\newtheorem{remark}[theorem]{Remark}
\newtheorem{remarks}[theorem]{Remarks}
\newtheorem{cor}[theorem]{Corollary}
\newtheorem{pro}[theorem]{Proposition}
\newtheorem{proposition}[theorem]{Proposition}
\newtheorem{exercise}[theorem]{Exercise}

\renewcommand{\theequation}{\thesection.\arabic{equation}}

%
% 1.) check labels and references 
%
% 2.) alles lesen und abschleifen 
%
% 3.) erst alles hintereinander aufschreiben, dann aufteilen 
%
% 4.) Lsgen exercises 
%
% 5.) $\beta$ vs $\ell$ 
%
% 6.) continuous vs essential spectrum 
%
% 7.) was ist Lemma, Thm., ... 
%
% 8.) Bez. static, steady state, ... 
% 
% 9.) zweiter Druckfehler 
%
% 10.) ueberarbeiten, Einleitung 
%
% 11.) ?? weg etc., auch bei citations 
%
% 12.)  
%
%
%
%

%%%%%%%%%%%%%%%%%%%%%%%%%%%%%%%%%%%%%%%%%%%%%%%%%%%%%%%%%%%%%%%%%%%%%%%%%%%%%%%

\title{A Birman-Schwinger principle in galactic dynamics
       \\[1ex] ESI, Vienna, Feb.~07-11, 2022}

\author{Markus Kunze\\
        Mathematisches Institut\\
        Universit\"at K\"oln\\
        Weyertal 86-90\\
        D-50931 K\"oln, Germany\\
        email: mkunze@math.uni-koeln.de}

\maketitle

%%%%%%%%%%%%%%%%%%%%%%%%%%%%%%%%%%%%%%%%%%%%%%%%%%%%%%%%%%%%%%%%%%%%%%%%%%%%%%%

\begin{abstract} These are the (somewhat extended) lecture notes for four lectures 
delivered at the spring school during the thematic programme ``Mathematical Perspectives 
of Gravitation beyond the Vacuum Regime'' at ESI Vienna in February 2022.  
\end{abstract}

%%%%%%%%%%%%%%%%%%%%%%%%%%%%%%%%%%%%%%%%%%%%%%%%%%%%%%%%%%%%%%%%%%%%%%%%%%%%%%%

\tableofcontents

%%%%%%%%%%%%%%%%%%%%%%%%%%%%%%%%%%%%%%%%%%%%%%%%%%%%%%%%%%%%%%%%%%%%%%%%%%%%%%%

\section{Introduction}
\setcounter{equation}{0}

These are the (somewhat extended) lecture notes for four lectures 
delivered at the spring school during the thematic programme ``Mathematical Perspectives 
of Gravitation beyond the Vacuum Regime'' at ESI Vienna in February 2022.  
The main reference for the lectures is \cite{MK}, which has some overlap 
with \cite{BAH}, although we wanted to emphasize the action-angle 
variables approach and put a main focus on the Birman-Schwinger principle, 
as is done in \cite{MK}. Since the lectures have been aimed 
at newcomers, some parts of them cover basic background material. 
The author is grateful to the organizers H.~Andr\'{e}asson, D.~Fajman, 
J.~Joudioux and T.~Oliynyk for making the programme happen despite the pandemic, 
and thanks are due to the ESI for providing a very stimulating working 
atmosphere. 

%%%%%%%%%%%%%%%%%%%%%%%%%%%%%%%%%%%%%%%%%%%%%%%%%%%%%%%%%%%%%%%%%%%%%%%%%%%%%%%

\section{The Birman-Schwinger principle in quantum mechanics} 
\setcounter{equation}{0}

The Birman-Schwinger principle is a widely used 
and well-established tool in mathematical quantum mechanics. 
It was introduced through the independent works of Birman \cite{bir} 
and Schwinger \cite{schwi}, 
with the idea of counting, or at least estimating, the number of eigenvalues 
of Schr\"odinger operators on $L^2(\R^n)$. To be more specific, consider 
(only formal at this point) 
\[ H=-\Delta+V; \]  
to avoid introducing negative parts we will assume that $V\le 0$. 

\begin{theorem}\label{BSinQM} The following assertions hold: 
\noindent  
\begin{itemize} 
\item[(a)] $-e$ is a (negative) eigenvalue of $H$ 
if and only if $1$ is an eigenvalue of the Birman-Schwinger operator 
\begin{equation}\label{Be} 
   B_e=\sqrt{-V}(-\Delta+e)^{-1}\sqrt{-V}. 
\end{equation} 
\end{itemize} 
Furthermore, 
\begin{itemize} 
\item[(b)] if $\phi$ is an eigenfunction of $H$ for the eigenvalue $-e$, 
then $\psi=\sqrt{-V}\phi$ is an eigenfunction of $B_e$ for the eigenvalue $1$; 
\item[(c)] if $\psi$ is an eigenfunction of $B_e$ for the eigenvalue $1$, 
then $\phi=(-\Delta+e)^{-1}(\sqrt{-V}\psi)$ is an eigenfunction of $H$ 
for the eigenvalue $-e$. 
\end{itemize} 
\end{theorem} 
{\bf Proof\,:} See \cite[Section 4.3.1]{LS}. 
If $H\phi=(-\Delta+V)\phi=(-e)\phi$, then we define $\psi=\sqrt{-V}\phi$ to obtain  
\[ B_e\psi=\sqrt{-V}(-\Delta+e)^{-1}\sqrt{-V}\sqrt{-V}\phi
   =\sqrt{-V}(-\Delta+e)^{-1}(-V)\phi
   =\sqrt{-V}\phi=\psi. \] 
Conversely, if $B_e\psi=\psi$ holds and if we put $\phi=(-\Delta+e)^{-1}(\sqrt{-V}\psi)$, 
it follows that 
\[ (-\Delta+e)\phi=\sqrt{-V}\psi
   =\sqrt{-V}\sqrt{-V}(-\Delta+e)^{-1}\sqrt{-V}\psi=(-V)\phi, \] 
and hence $H\phi=(-\Delta+V)\phi=(-e)\phi$, which completes the argument. 
{\hfill$\Box$}\bigskip  

Here are some facts: 

\begin{itemize} 
\item The operators $B_e$ are non-negative Hilbert-Schmidt operators 
(if $V$ decays sufficiently fast and $n\le 3$), 
and in particular they are compact. 
\item Their eigenvalues can be ordered: $\lambda_1(e)\ge\lambda_2(e)\ge\ldots\to 0$ 
and the eigenvalue curves are decreasing in $e$, in that $\tilde{e}\ge e$ implies that 
$\lambda_k(e)\le\lambda_k(\tilde{e})$ for all $k$. 
\item The number of eigenvalues of $H$ 
less than or equal to $-e$ agrees with the number of eigenvalues of $B_e$ greater than 
or equal to $1$, counting multiplicities in both cases; cf.~\cite[Figure 4.1, p.~78]{LS} 
for an illustration. 
\item Not only the number of eigenvalues of $H$ can be bounded, 
but for instance also eigenvalue moments like $\sum_j |-e_j|^\gamma$, 
where the sum extends over all 
negative eigenvalues $-e_j$ of $H$. This fact lies at the heart 
of many important results in the field. 
Let us only mention here the Lieb-Thirring bound 
\[ \sum_j |-e_j|\le L_{1, 3}\int_{\R^3} |V(x)|^{5/2}\,dx \] 
in three dimensions for an absolute constant $L_{1, 3}>0$ and $V\in L^{5/2}(\R^3)$. 
It is used in those authors' proof of the stability of matter \cite{LT}, 
which has found many generalizations \cite{LS}, and which is much easier to follow 
than the original argument by Dyson and Lenard \cite{DL}. 
\end{itemize} 

\noindent
Good general textbooks that cover the Birman-Schwinger principle 
are \cite[Section 4.3]{LS}, \cite{RSIV,Sim_quadr,FunctInt} or \cite[Section 7.9]{Simon4}. 

There is also a large number of further applications of the Birman-Schwinger principle: 

\begin{itemize}
\item complex-valued potentials 
\item Dirac operators
\item Bardeen-Cooper-Schrieffer model of superconductivity 
\item linearized 2D Euler equations 
\item ... and many more. 
\end{itemize}  

%%%%%%%%%%%%%%%%%%%%%%%%%%%%%%%%%%%%%%%%%%%%%%%%%%%%%%%%%%%%%%%%%%%%%%%%%%%%%%%

\section{Galactic dynamics: The Vlasov-Poisson system} 
\setcounter{equation}{0}

\begin{itemize}
\item Galactic dynamics in general refers to the modeling of the time evolution 
of self-gravitating matter 
such as galaxies, or on an even larger scale, clusters of galaxies. 
\item $N$-body problem, with $N$ quite large: $N\sim 10^6-10^{11}$ for galaxies 
and $N\sim 10^2-10^3$ for clusters of galaxies. 
\item The $N$-body problem consists of coupled Newtonian equations, 
one for each individual object 
(the `objects' in a galaxy are stars, 
those in a cluster of galaxies are galaxies), and to study the collective 
behavior of the system. 
\item While results may be obtainable numerically in this way, 
the mathematical complexity of even 
the three-body problem prevents one from rigorously addressing deeper questions 
(concerning for instance galaxy formation or stability) for such stellar systems. 
\item From the early days of the field, a statistical description of the evolution 
has been proposed, by Vlasov \cite{Vlasov_paper} in 1938 for plasmas 
(in this case a related equation is satisfied) 
and by Jeans \cite{Jeans_paper} in 1915 for gravitational systems; see \cite{Henon} 
for an interesting historical discussion of the origins of the equation. 
It is also known as the `collisionless Boltzmann equation', 
which refers to the fact that collisions among the stars or galaxies are sufficiently rare 
to be neglected. A standard source of information on galactic dynamics is \cite{BT}. 
\end{itemize} 

The time evolution of such a system is then governed by a distribution function 
$f=f(t, x, v)$ that depends on time $t\in\R$, position $x\in\R^3$ and velocity $v\in\R^3$. 
The quantity $\int_{{\cal X}} dx\int_{{\cal V}} dv\,f(t, x, v)$ should be thought 
of as the number of objects 
(henceforth called `particles') at time $t$, which are located at some point 
$x\in {\cal X}\subset\R^3$ 
and which have velocities $v\in {\cal V}\subset\R^3$. Each individual particle 
follows a trajectory $(X(s), V(s))$ in phase space $\R^3\times\R^3$ 
such that $(X(t), V(t))=(x, v)$ at time $t$ and 
\begin{equation}\label{charglen} 
   \dot{X}(s)=V(s),\quad\dot{V}(s)=-\nabla_x U(s, X(s)),
\end{equation} 
where $F=-\nabla_x U$ denotes the Coulomb-type force field that 
is collectively generated by all particles. 
The requirement that $f$ be constant along the curves defined by (\ref{charglen}) 
then leads to the relation 
\begin{eqnarray}\label{riqu} 
   0 & = & \frac{d}{ds}\,[f(s, X(s), V(s))] 
   \nonumber
   \\ & = & \partial_t f(s, X(s), V(s))+V(s)\cdot\nabla_x f(s, X(s), V(s))
   -\nabla_x U(s, X(s))\cdot\nabla_v f(s, X(s), V(s))
   \nonumber \\ & & 
\end{eqnarray} 
for all $s$. Evaluated at time $t$, this yields 
\begin{equation}\label{vpgr1} 
   \partial_t f(t, x, v)+v\cdot\nabla_x f(t, x, v)-\nabla_x U_f(t, x)\cdot
   \nabla_v f(t, x, v)=0 
\end{equation} 
for all $(t, x, v)$, which is usually called the Vlasov equation 
(despite the historic inadequacy of this terminology). 
The next step is to express the force field $F$ in terms of the distribution function $f$. 
Since we are aiming at describing gravitational binding, we need to have 
$F\sim -\nabla_x V_{{\rm C}}$ 
for the Coulomb potential $V_{{\rm C}}(x)=-\frac{1}{|x|}$ at large distances. 
This suggests to use the field $F=-\nabla_x U$ 
induced by the Poisson equation 
\begin{equation}\label{vpgr2} 
   \Delta_x U_f(t, x)=4\pi\rho_f(t, x),\quad\lim_{|x|\to\infty} U_f(t, x)=0,
   \quad\mbox{where}\quad\rho_f(t, x)=\int_{\R^3} f(t, x, v)\,dv
\end{equation} 
denotes the charge density induced by $f$. Observe that $\int_{{\cal X}} dx\,\rho_f(t, x)$ 
represents the number of particles at time $t$, of any velocity, 
which are located at some point $x\in {\cal X}$. Then 
\begin{equation}\label{vpgr2b} 
   U_f(t, x)=-\int_{\R^3}\frac{\rho_f(t, y)}{|y-x|}\,dy.
\end{equation} 
is Coulomb-like as $|x|\to\infty$. 

Initial data $f(0, x, v)=f_0(x, v)$ 
at time $t=0$ have to be specified for $f$ only, since then (\ref{vpgr2b}) 
determines the initial data $U_f(0, x)$. We will exclusively be interested 
in classical solutions 
of (\ref{vpgr1}), (\ref{vpgr2}), whose global-in-time existence is ensured, 
under reasonable assumptions on $f_0$. 

\begin{theorem}[\cite{pfaff,schaeff} and \cite{LP}] 
Let $f_0$ be continuously differentiable and compactly supported. 
Then the Vlasov-Poisson system (\ref{vpgr1}), (\ref{vpgr2}), (\ref{vpgr2b}) 
has a global and unique solution. 
\end{theorem} 

For a mathematical overview of the system and more background material 
the reader may wish to consult \cite{Glassey_book,Mouhot_rev,Reinrev}. 
Throughout the course we will adopt a dynamical systems viewpoint: 
Given some initial data $f_0$ we are interested in what happens to the resulting 
solution $f(t)$ (that lies in a space of functions depending on $(x, v)$) 
as $t\to\infty$? 

\section{Spherically symmetric solutions} 
\setcounter{equation}{0}

Almost exclusively we will be dealing with spherically symmetric 
solutions of the Vlasov-Poisson system. 
A function $g=g(x, v)$ is said to be spherically symmetric, 
if $g(Ax, Av)=g(x, v)$ for all $A\in {\rm SO}(3)$ and $x, v\in\R^3$.
Expressed in more sophisticated terms, $g$ needs to be equivariant 
w.r.~to the group action ${\rm SO}(3)\times (\R^3\times\R^3)\to\R^3\times\R^3$, 
$(A, x, v)\mapsto (Ax, Av)$. In this case $\rho_g(x)=\rho_g(r)$ 
and $U_g(x)=U_g(r)$ are radially symmetric; here $r=|x|$. More explicitly, 
\begin{eqnarray}
   U_g(r) & = & -\frac{4\pi}{r}\int_0^r s^2\rho_g(s)\,ds-4\pi\int_r^\infty s\rho_g(s)\,ds,
   \label{Ugsymm}  
   \\ U'_g(r) & = & \frac{4\pi}{r^2}\int_0^r s^2\rho_g(s)\,ds
   =\frac{1}{r^2}\int_{|x|\le r}\rho_g(x)\,dx,
   \nonumber 
\end{eqnarray} 
where $'$ denotes $\frac{d}{dr}$. 

\begin{exercise} Prove (\ref{Ugsymm}) from (\ref{vpgr2b}). 
\end{exercise} 

It can be shown that a spherically symmetric function $g=g(x, v)$ 
does in fact only depend upon three variables: $g=\tilde{g}(|x|, |v|, x\cdot v)$.

\begin{exercise} Establish this claim. 
\end{exercise} 

In spherical symmetry, the variables 
\[ r=|x|,\quad p_r=\frac{x\cdot v}{r}\quad\mbox{and}\quad L=x\wedge v, \] 
are most useful. Here $p_r\in\R$ denotes the radial momentum and $L\in\R^3$ 
is the angular momentum. Since 
\begin{equation}\label{amdef} 
   |L|^2=|x|^2|v|^2-(x\cdot v)^2=r^2(|v|^2-p_r^2),
\end{equation} 
we get $|v|^2=\frac{\ell^2}{r^2}+p_r^2$ for $\ell=|L|$. 
This implies that a function $g=\tilde{g}(|x|, |v|, x\cdot v)$ 
can equally well be expressed as a function $g=\hat{g}(r, p_r, \ell)$; 
of course we are going to identify all versions of $g$. 

If we restrict to spherically symmetric solutions, i.e., 
$f(t)=f(t, \cdot, \cdot)$ is spherically symmetric for all $t$, 
then in the new variables $(r, p_r, \ell)$ the Vlasov-Poisson system 
can be rewritten as 
\[ \partial_t f(t, r, p_r, \ell^2)+p_r\,\partial_r f(t, r, p_r, \ell^2)
   +\Big(\frac{\ell^2}{r^3}-\partial_r U_f(t, r)\Big)\,\partial_{p_r} 
   f(t, r, p_r, \ell^2)=0 \] 
and 
\[ U''_f(t, r)+\frac{2}{r}\,U'_f(t, r)=4\pi\rho_f(t, r),
   \quad\lim_{r\to\infty} U_f(t, r)=0,
   \quad\rho_f(t, r)=\frac{2\pi}{r^2}\int_0^\infty d\ell\,\ell
   \int_{\R} dp_r f(t, r, p_r, \ell^2), \] 
and moreover 
\begin{equation}\label{hari} 
   U_f(t, r) = -\frac{4\pi}{r}\int_0^r\sigma^2\rho_f(t, \sigma)\,d\sigma
   -4\pi\int_r^\infty s\rho_f(t, \sigma)\,d\sigma
\end{equation}  
due to (\ref{Ugsymm}). From (\ref{hari}) it is plain to see that 
$U_f(t, r)\sim -\frac{M}{r}$ as $r\to\infty$ 
for $M=4\pi\int_0^\infty \sigma^2\rho_f(t, \sigma)\,d\sigma
=\int_{\R^3}\int_{\R^3} f(t, x, v)\,dx\,dv$ denotes the mass 
of the system.  

%%%%%%%%%%%%%%%%%%%%%%%%%%%%%%%%%%%%%%%%%%%%%%%%%%%%%%%%%%%%%%%%%%%%%%%%%%%%%%%

\section{Steady state solutions} 
\setcounter{equation}{0}

From a dynamical systems perspective, the easiest solutions 
of dynamical relevance are steady states, i.e., time-independent solutions. 
The Vlasov-Poisson system possesses an abundance of such solutions $Q=Q(x, v)$, 
which we seek to be spherically symmetric. 
Let $e_Q(x, v)=\frac{1}{2}\,|v|^2+U_Q(x)$ denote the particle energy 
and let $\ell^2=|L|^2$ be as in (\ref{amdef}).  

\begin{lemma}\label{el2}  
Both $e_Q$ and $\ell^2$ are conserved along solutions 
of the characteristic equation $\ddot{X}(t)=-\nabla U_Q(X(t))$ from (\ref{charglen}). 
\end{lemma} 
{\bf Proof\,:} Let us consider $\ell^2$ for example. Then for $V=\dot{X}$ 
\begin{eqnarray*}
   \frac{d}{dt}\,|X(t)\wedge V(t)|^2
   & = & 2(X(t)\wedge V(t))\cdot\Big(\dot{X}(t)\wedge V(t)+X(t)\wedge\dot{V}(t)\Big)
   \\ & = & -2(X(t)\wedge V(t))\cdot\Big(X(t)\wedge\nabla U_Q(X(t))\Big).
\end{eqnarray*} 
From $U_Q(x)=U_Q(|x|)=U_Q(r)$ we deduce $\nabla U_Q(x)=U'_Q(r)\frac{x}{|x|}$, 
and thus $X(t)\wedge\nabla U_Q(X(t))=0$. 
Note this argument has nothing to do with Vlasov-Poisson, 
but only relies on the fact that $U_Q$ is what is called a central potential. 
The calculation for $e_Q$ is similar. 
{\hfill$\Box$}\bigskip  

Lemma \ref{el2} is the key to obtaining steady state solutions: 
If we seek a solution in the form $Q(x, v)=Q(e_Q, \ell^2)$ 
(observe the abuse of notation here), then (\ref{riqu}), 
i.e., the Vlasov equation will automatically be satisfied. 
Thus finding a steady state comes down to solving the semilinear equation
\begin{equation}\label{Uhugl} 
   \frac{1}{r^2}(r^2 U'(r))'=\Delta U(x)
   =4\pi\int_{\R^3}Q\Big(\frac{1}{2}\,|v|^2+U(x), |x\wedge v|^2\Big)\,dv,
   \quad\lim_{r\to\infty} U(r)=0,
\end{equation} 
for $U=U_Q$, if the profile function $Q=Q(e_Q, \ell^2)$ is given.  
In fact it is the content of Jeans's theorem that the distribution function $Q$ 
of every spherically symmetric steady state solution has to be of the form 
$Q=Q(e_Q, \ell^2)$. 

Therefore the question arises for which $Q$'s (\ref{Uhugl}) can be solved, 
and it turns out that there is a variety of possible choices, even if we restrict ourselves 
to the easier case that $Q=Q(e_Q)$ does not depend on $\ell^2$; 
such steady state solutions are called isotropic. There is a vast literature 
concerning different classes of ansatz functions (called polytropes, King models, ... etc.), 
see \cite{BT,Mouhot_rev,Reinrev}, but for the purpose of this course 
it will be sufficient to keep in mind the example of the polytropes. 
They are given by 
\begin{equation}\label{poly_form} 
   Q(e_Q)=(e_0-e_Q)_+^k
\end{equation} 
for a fixed cut-off energy $e_0<0$ and $k\in ]-\frac{1}{2}, \frac{7}{2}[$; 
here $s_+=\max\{s, 0\}$. Then  
\begin{equation}\label{rhopoly} 
   \rho_Q(r)=c_n(e_0-U_Q(r))_+^n,\quad n=k+\frac{3}{2}\in ]1, 5[,
   \quad c_n=(2\pi)^{3/2}\,\frac{\Gamma(k+1)}{\Gamma(k+\frac{5}{2})}.
\end{equation} 
   
\begin{exercise} Prove (\ref{rhopoly}). 
\end{exercise} 
   
The potential $U_Q$ is not known explicitly. 
All the polytropic steady state solutions do have finite radius $r_Q$ (i.e., 
the density $\rho_Q$ is supported in $[0, r_Q]$ and $r_Q<\infty$) and finite mass 
$M_Q=\int_{\R^3}\rho_Q(x)\,dx=4\pi\int_0^{r_Q} r^2\rho_Q(r)\,dr=\int_{\R^3}
\int_{\R^3} Q(x, v)\,dx\,dv$. 
The limiting case $k=7/2$ is called the Plummer sphere, where $M_Q$ is still finite, 
but $r_Q=\infty$.  
It is also important to note that $Q'(e_Q)<0$ inside the support of the polytropes. 
In general, this property is very much tied to (linear) stability.     
 
%%%%%%%%%%%%%%%%%%%%%%%%%%%%%%%%%%%%%%%%%%%%%%%%%%%%%%%%%%%%%%%%%%%%%%%%%%%%%%% 
 
\section{Action angle variables} 
\setcounter{equation}{0}

Action angle variables are particularly well-suited for Hamiltonian systems. 
We start with a one-degree-of freedom example, a reliable general source 
on the topic being \cite{zehnder}. 

\begin{example}[Action-angle variables]
{\rm We consider $n=1$, $q, p\in\R$ and
\[ H(q, p)=\frac{1}{2}\,p^2+V(q), \]
where the potential $V$ should be such that the phase portrait of the resulting system
$\ddot{q}=-V'(q)$ contains a fixed point (which we take to be the origin) that is encircled
by a family of periodic solutions $\gamma_h$ which is parameterized by their energy $h$
in some interval $h\in ]h_0, h_1[$. Then
\[ \gamma_h\subset\Big\{(q, p)\in\R\times\R: \frac{1}{2}\,p^2+V(q)=h\Big\}, \] 
but not necessarily $\gamma_h=\{\ldots\}$, since the energy level set $\{\ldots\}$
may consist of several components; for instance this is the case for
$V(q)=q^2 (q^2-1)$, or a version thereof shifted appropriately to center one set
of periodic orbits at the origin, where the left and the right interior
of the homoclinic orbits both contain solutions of the same period and energy.
Let $q_{\pm}(h)$ denote the intersections of $\gamma_h$ and the $q$-axis $\{p=0\}$,
i.e., $V(q_{\pm}(h))=h$ is required along with $q_-(h)<0<q_+(h)$. If $T(h)$ is the period
of $\gamma_h$, then
\begin{equation}\label{th-allg}
   T(h)=2\int_{q_-(h)}^{q_+(h)}\,\frac{ds}{\sqrt{2(h-V(s))}}.
\end{equation}
Next denote by $2\pi I(h)$ the area encircled by $\gamma_h$. Since the orbit
is transversed in the clockwise direction, the Green-Riemann formula says that
\[ 2\pi I(h)=\int_{\gamma_h} p\,dq \]
for the action $I$. Furthermore, elementary calculus tells us that
\[ 2\pi I(h)=2\int_{q_-(h)}^{q_+(h)}\sqrt{2(h-V(s))}\,ds, \] 
noting that the height function at $q$ is just $\pm p=\sqrt{2(h-V(q))}$.
Recalling that $V(q_{\pm}(h))=h$, we see that in particular
\[ I'(h)=\frac{1}{2\pi}\,T(h)>0 \] 
holds. Thus the function $h\mapsto I(h)$ (on $]h_0, h_1[$)
admits an inverse function that is denoted by $I\mapsto h(I)$.
Differentiating the relation $h(I(h))=h$, we get
\[ 1=h'(I(h))I'(h)=\frac{1}{2\pi}\,h'(I(h))\,T(h). \]

Now we consider the transformation
\begin{equation}\label{howos}
   \Phi: (q, p)\mapsto (\theta, I)
\end{equation}
that is obtained from the so--called generating function
\[ S(q, I)=\int_{q_-(h(I))}^q\sqrt{2(h(I)-V(s))}\,ds. \]
In general, a generating function depends on one 
``old'' variable (here: $q$) and one ``new'' variable (here: $I$). 
This means that $p=\partial_q S$ and $\theta=\partial_I S$, 
in the following sense: Given $(q, p)$,
where for instance $p\ge 0$, the relation
\begin{equation}\label{genfct-p}
   p=\partial_q S(q, I)=\sqrt{2(h(I)-V(q))}
\end{equation}
has to be solved for $I=I(q, p)$, and then the assignment
\[ \theta(q, p)=\partial_I S(q, I(q, p)) \]
completes the definition of the transformation (\ref{howos}).

A key feature of transformations derived from generating functions 
in this way is that they are canonical (i.e., symplectic).  
To see this, differentiating $p=\partial_q S(q, I)$ w.r.~to $p$ 
implies that $1=(\partial^2_{qI}S)(\partial_p I)$. 
Therefore we deduce from $\theta=\partial_I S(q, I)$ that 
\begin{eqnarray*} 
   d\theta\wedge dI & = & \Big[(\partial_{qI}^2 S)\,dq+(\partial_{II}^2 S)\,dI\Big]\wedge dI
   =(\partial_{qI}^2 S)\,dq\wedge dI
   \\ & = & (\partial_{qI}^2 S)\,dq\wedge\Big[(\partial_q I)\,dq+(\partial_p I)\,dp_r\Big]
   =(\partial_{qI}^2 S)(\partial_p I)\,dq\wedge dp=dq\wedge dp, 
\end{eqnarray*} 
which means that $\Phi$ from (\ref{howos}) is indeed canonical. 

The meaning of the angular variable $\theta$ is as follows. Denote by
\[ \tau(q, p)=\int_{q_-(h)}^q\,\frac{ds}{\sqrt{2(h-V(s))}} \]
the time that it takes the solution, if for instance $p>0$,
to pass from $(q_-(h), 0)$ to $(q, p)$ on $\gamma_h$. Noting that
\[ \theta(q, p)=\partial_I S(q, I)
   =\int_{q_-(h)}^q\,\frac{ds}{\sqrt{2(h(I)-V(s))}}\,h'(I)
   =\tau(q, p)\,\frac{2\pi}{T(h)},  \] 
this can be rewritten to read
\[ \frac{\theta(q, p)}{2\pi}=\frac{\tau(q, p)}{T(h)}. \]
Hence $\theta(q, p)\in [0, 2\pi[$ is the angle of clockwise rotation
of the line segment $[(q_-(h), 0), (0, 0)]$
to the line segment $[(q, p), (0, 0)]$. The variables $I$ and $\theta$
are called action and angle variables, respectively.

Since the transformation is canonical, it is sufficient to transform 
the Hamiltonian function to obtain the equations of motion 
in the new variables. In this case we obtain 
\[ {\cal H}(\theta, I)=H(\Phi^{-1}(\theta, I))
   =\frac{1}{2}\,p^2+V(q)=h(I) \]
by (\ref{genfct-p}). Here we see the main reason for passing to action-angle variables:
the dynamics become very simple, since in the new variables the Hamiltonian 
is independent of the angular variable. The associated equations of motion are
\[ \dot{\theta}=\partial_I {\cal H}=h'(I)=:\omega(I),
   \quad\dot{I}=-\partial_\theta {\cal H}=0, \]
and the corresponding solutions are
\[ \theta(t)=\theta_0+\omega(I_0)t,\quad I(t)=I_0, \] 
which is an angular rotation with frequency $\omega(I_0)$.
{\hfill$\diamondsuit$}
}
\end{example}
\medskip

\begin{exercise} Prove the period relation \ref{th-allg}. 
\end{exercise} 

\begin{exercise} 
Let $V(q)=\frac{\omega^2}{2}\,q^2$, i.e., we consider the harmonic oscillator 
$\ddot{q}+\omega q=0$ with mass $m=1$. Show the following items: 
\begin{itemize}
\item[(a)] The intersection points of the orbit of energy $h$ with the $q$-axis 
are $q_{\pm}(h)=\pm\frac{\sqrt{2h}}{\omega}$. 
\item[(b)] The period function is $T(h)=\frac{2\pi}{\omega}$, independently of $h$. 
\item[(c)] One has $I(h)=\frac{1}{\omega}\,h$ and $I'(h)=\frac{1}{\omega}
=\frac{1}{2\pi}\,T(h)$. 
The inverse function to $h\mapsto I(h)$ is $h(I)=\omega I$ which yields 
the constant frequency $\omega(I)=h'(I)=\omega$. 
\item[(d)] The generating function is 
\[ S(q, I)=2I\int_{-1}^{\sqrt{\frac{\omega}{2I}}\,q}\sqrt{1-\tau^2}\,d\tau \] 
(and there is no need to evaluate the integral explicitly). 
\item[(e)] Calculate that $\partial_I S(q, I(q, p))=\theta$ and 
\[ \Phi^{-1}(\theta, I)=\bigg(-\sqrt{\frac{2I}{\omega}}\,\cos\theta, \sqrt{2I\omega}
   \,\sin\theta\bigg). \]  
\item[(f)] Establish that 
\[ q(t)=-\sqrt{\frac{2I_0}{\omega}}\,\cos(\theta_0+\omega t),\quad p(t)
   =\sqrt{2I_0\omega}\,\sin (\theta_0+\omega t), \] 
is the solution such that $\Phi(q_0, p_0)=(\theta_0, I_0)$. 
\end{itemize} 
\end{exercise}

Now we return to the Vlasov-Poisson setting and consider the characteristic equation 
\begin{equation}\label{barol} 
   \ddot{X}=-\nabla U_Q(X(t))
\end{equation} 
for an isotropic steady state solution $Q=Q(e_Q)$; 
it is (an autonomous) Hamiltonian system. 
By the spherical symmetry, one can use a canonical change of variables
\begin{equation}\label{1stchange} 
   (x, v)\mapsto (p_r, L_3, \ell; r, \varphi, \chi)
\end{equation} 
on the support $K={\rm supp}\,Q$ of $Q$ as described in \cite[Ch.~3.5.2]{BT} 
and \cite[\S\,5.3]{thirr} to simplify matters considerably. 
Let us first have a look at the variables on the right-hand side of (\ref{1stchange}). 
Since $L=x\wedge v$, we have $L_3=x_1v_2-x_2v_1$ for the third component. 
The angles $\varphi$ and $\chi$ are determined by 
\begin{eqnarray*} 
   & & \sin\varphi=\frac{L_1}{(\ell^2-L_3^2)^{1/2}},\quad\cos\varphi
   =\frac{L_2}{(\ell^2-L_3^2)^{1/2}}, 
   \\ & & \cos\chi=\frac{(e_3\wedge L)\cdot x}{r(\ell^2-L_3^2)^{1/2}},
   \quad\sin\chi=\frac{\ell\,x_3}{r(\ell^2-L_3^2)^{1/2}}.
\end{eqnarray*} 
From these relations it can be calculated that indeed (\ref{1stchange}) is canonical. 
The variable pairs $r\leftrightarrow p_r$, $\varphi\leftrightarrow L_3$, 
and $\chi\leftrightarrow\ell$ are conjugate, their Poisson brackets 
can be evaluated explicitly; 
see \cite[\S\,5.3]{thirr}, also for an illustration of how the new coordinates 
can be read off. 
The Hamiltonian function for (\ref{barol}) is $e_Q(x, v)=\frac{1}{2}\,|v|^2+U_Q(x)$. 
Since the transformation is canonical, we only need to transform the Hamiltonian 
in order to obtain (\ref{barol}) in the new coordinates. 
Recalling that $|v|^2=\frac{\ell^2}{r^2}+p_r^2$, it is found that 
\[ e_Q(r, p_r, \ell)=\frac{1}{2}\,p_r^2+U_{{\rm eff}}(r, \ell),
   \quad\mbox{with}\quad U_{{\rm eff}}(r, \ell)=U_Q(r)+\frac{\ell^2}{2r^2} \] 
being the effective potential. Now 
\[ \dot{r}=\frac{\partial e_Q}{\partial p_r}=p_r,
   \quad\dot{p}_r=-\frac{\partial e_Q}{\partial r}=-U'_{{\rm eff}}(r, \ell), \] 
thus the resulting equation of motion is 
\[ \ddot{r}=-U'_{{\rm eff}}(r, \ell). \] 
This should be viewed as one second order Hamiltonian system in $(r, p_r)$ 
per each fixed $\ell$, where the potential is given by $r\mapsto U_{{\rm eff}}(r, \ell)$. 
The potential has the following shape: 

\begin{figure}[H]
   \centering
   \includegraphics[angle=0,width=0.5\linewidth]{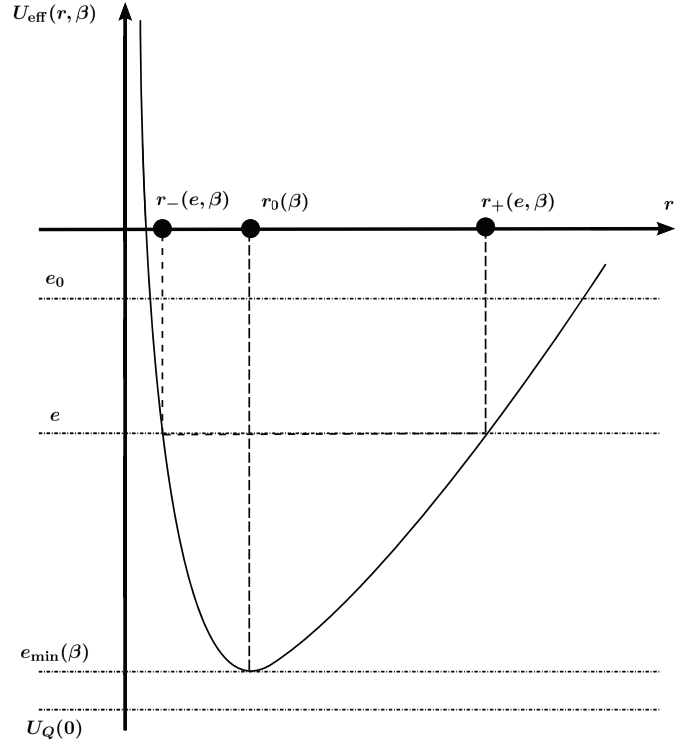}
   % \\[-10ex]
   \caption{The effective potential $U_{{\rm eff}}(r, \ell)=U_{{\rm eff}}(r, \beta)$}
\end{figure}

\noindent
There are exactly two zeros $0<r_-(e, \ell)<r_+(e, \ell)$ of $0=2(e-U_{{\rm eff}}(r, \ell))$ 
and the minimum is attained at a unique point $r_0(\ell)$. 

The new variables $(p_r, L_3, \ell; r, \varphi, \chi)$ in (\ref{1stchange}) 
are not yet the desired action-angle variables, since $e=e(r, p_r, \ell)$ depends upon $r$, 
which plays the role of an angle; remember from the 1D example above 
that the goal is to get the Hamiltonian independent of the angle(s). 
Therefore a further canonical transformation 
\begin{equation}\label{WWV} 
   (r, p_r)\to (\theta, I)\quad\mbox{at a fixed}\,\,\ell
\end{equation} 
will be made. At such a fixed $\ell$, we can do this 
in a region where the orbits of $U_{{\rm eff}}(\cdot, \ell)$ are periodic; 
it is achieved by means of a generating function as above. 
The angle $\theta\in [0, \pi]$ corresponds to one half-turn of the periodic orbit $\gamma$ 
in the potential $U_{{\rm eff}}(\cdot, \ell)$, connecting the `pericenter' $r_-$ 
to the `apocenter' $r_+$; here $\dot{r}=p_r>0$ for $r\in ]r_-, r_+[$ 
and $p_r(r_{\pm})=0$. Therefore if $\theta\in [\pi, 2\pi]$, then 
\begin{equation}\label{rprsymm} 
   r(\theta, I, \ell)=r(2\pi-\theta, I, \ell)
   \quad\mbox{and}\quad p_r(\theta, I, \ell)=-p_r(2\pi-\theta, I, \ell).
\end{equation} 
In other words, we need to determine the (inverse) transformation 
$(\theta, I)\mapsto (r, p_r)$ 
only for $\theta\in [0, \pi]$, where we have $p_r\ge 0$. 
Let $E=E(I, \ell)$ be the solution to 
\[ I=\frac{1}{2\pi}\int_\gamma p_r\,dr
   =\frac{1}{\pi}\int_{r_-(E,\,\ell)}^{r_+(E,\,\ell)}
   \sqrt{2(E-U_{{\rm eff}}(r, \ell))}\,dr, \] 
where $\gamma$ is as before. Then consider
\begin{equation}\label{grieg} 
   S(r, I, \ell)=\int_{r_-(E(I,\,\ell),\,\ell)}^r
   \sqrt{2(E(I, \ell)-U_{{\rm eff}}(r', \ell))}\,dr'
\end{equation} 
as a generating function for (\ref{WWV}). The rules for determining the full transformation 
from $S$ are once again given by 
\[ \theta=\partial_I S,\quad p_r=\partial_r S. \] 
More precisely, the equation 
\begin{equation}\label{thetagl} 
   \theta=\partial_I S(r, I, \ell)
\end{equation} 
has a solution $r=r(\theta, I, \ell)$. In addition, put 
\[ p_r=p_r(\theta, I, \ell)=\partial_r S(r(\theta, I, \ell), I, \ell). \] 
Thus more explicitly 
\[ p_r(\theta, I, \ell)=\sqrt{2(E(I, \ell)-U_{{\rm eff}}(r(\theta, I, \ell), \ell))}, \] 
which yields 
\[ E(I, \ell)=\frac{1}{2}\,p_r(\theta, I, \ell)^2+U_{{\rm eff}}(r(\theta, I, \ell), \ell)
   =e(r(\theta, I, \ell), p_r(\theta, I, \ell), \ell). \] 

Now $p_r=\partial_r S$ and (\ref{grieg}) yield $e=\frac{p_r^2}{2}
+U_{{\rm eff}}(r, \ell)=E(I, \ell)$, 
so $E$ will only depend upon action variables after the transformation (\ref{WWV}), 
which leads to the overall transformation 
\begin{equation}\label{oacan} 
   (x, v)\mapsto (p_r, L_3, \ell; r, \varphi, \chi)
   \mapsto (I, L_3, \ell; \theta, \varphi, \chi), 
\end{equation} 
cf.~(\ref{1stchange}). Hence after applying (\ref{oacan}) the particle energy 
does only depend upon $I$ and $\ell$, both of which are actions. 
The associated frequencies are 
\begin{equation}\label{om123} 
   \omega_1(I, \ell)=\frac{\partial E(I, \ell)}{\partial I},
   \quad\omega_2(I, \ell)=\frac{\partial E(I, \ell)}{\partial L_3}=0, 
   \quad\omega_3(I, \ell)=\frac{\partial E(I, \ell)}{\partial\ell},
\end{equation} 
and the period functions are 
\[ T_1(I, \ell)=\frac{2\pi}{\omega_1(I, \ell)},
   \quad T_3(I, \ell)=\frac{2\pi}{\omega_3(I, \ell)}. \]  
Also (\ref{thetagl}) yields 
\[ \theta=\partial_I S(r, I, \ell)=\omega_1(I, \ell)
   \int_{r_-(E(I,\,\ell),\,\ell)}^r\frac{dr'}
   {\sqrt{2(E(I, \ell)-U_{{\rm eff}}(r', \ell))}}. \] 
Since $\theta=0$ at $r_-$ and $\theta=\pi$ at $r_+$ (recall that $\dot{r}=p_r>0$ 
along this part of the orbit), we obtain 
\[ \pi=\frac{2\pi}{T_1(I, \ell)}
   \int_{r_-(E(I,\,\ell),\,\ell)}^{r_+(E(I,\,\ell),\,\ell)}
   \frac{dr}{\sqrt{2(E(I, \ell)-U_{{\rm eff}}(r, \ell))}}, \] 
or explicitly
\begin{equation}\label{T1form} 
   T_1(I, \ell)=2\int_{r_-(E(I,\,\ell),\,\ell)}^{r_+(E(I,\,\ell),\,\ell)}
   \frac{dr}{\sqrt{2(E(I, \ell)-U_{{\rm eff}}(r, \ell))}}
\end{equation} 
for the period function. In particular, $T_1(I, \ell)=T_1(E, \ell)$ 
by abuse of notation.  
\medskip 

To summarize, spherically symmetric functions $g=g(x, v)=g(|x|, |v|, x\cdot v)$ 
may also be expressed as $g=g(r, p_r, \ell)=g(\theta, I, \ell)$.

%%%%%%%%%%%%%%%%%%%%%%%%%%%%%%%%%%%%%%%%%%%%%%%%%%%%%%%%%%%%%%%%%%%%%%%%%%%%%%%

\section{Function spaces} 
\setcounter{equation}{0}

Next we consider the question of how $K={\rm supp}\,Q$ can be expressed in terms 
of the variables $\beta=\ell^2$ and $e=e_Q$. We will stick to the example 
of the polytropes (\ref{poly_form}), where 
\[ K=\{e_0-e_Q\ge 0\}. \] 
More precisely, since always $\theta\in [0, 2\pi]$ 
on $K$ for the angular variable $\theta$, we have to exhibit a set $D$ of $(e, \beta)$ 
such that 
\begin{equation}\label{zkm} 
   K=[0, 2\pi]\times D.
\end{equation} 
On this domain $D$ we need to have 
\begin{equation}\label{bladec} 
   e_0\ge e\ge U_{{\rm eff}}(r, \beta)\ge U_{{\rm eff}}(r_0(\beta), \beta)
   =U_Q(r_0(\beta))+\frac{\beta}{2r_0(\beta)^2},
\end{equation} 
with $r_0(\beta)$ denoting the unique point where the effective potential 
$U_{{\rm eff}}(r, \beta)=U_Q(r)+\frac{\beta}{2r^2}$ attains its minimum value 
$e_{{\rm min}}(\beta)=U_{{\rm eff}}(r_0(\beta), \beta)$. From (\ref{bladec}) we get 
\[ 2r_0(\beta)^2\,(e_0-U_Q(r_0(\beta))\ge\beta. \] 
Let 
\[ J=\{\beta\ge 0: 2r_0(\beta)^2\,(e_0-U_Q(r_0(\beta))\ge\beta\}. \] 

\begin{exercise} Prove that $J$ is an interval. You may use the general fact that 
$\beta\mapsto e_{{\rm min}}(\beta)$ is increasing. 
\end{exercise} 
{\bf Solution\,:} To see this, note that 
\[ 2r^2(e_0-U_{{\rm eff}}(r, \beta))+\beta=2r^2\Big(e_0-U_Q(r)-\frac{\beta}{2r^2}\Big)+\beta
   =2r^2(e_0-U_Q(r)). \] 
Therefore 
\[ 2r^2(e_0-U_Q(r))\ge\beta\quad\Longleftrightarrow\quad U_{{\rm eff}}(r, \beta)\le e_0, \] 
which implies that 
\begin{equation}\label{Jdom} 
   J=\{\beta\ge 0: e_{{\rm min}}(\beta)\le e_0\}.
\end{equation} 
Now $\beta\mapsto e_{{\rm min}}(\beta)$ is increasing by \cite[Lemma A.7(c)]{MK},  
and thus $J$ has to be an interval. 
{\hfill$\Box$}\bigskip  

\begin{exercise} Prove that $[0, \eps]\subset J$ for some $\eps>0$ small enough. 
You may use the general fact that $r_0(\beta)^4=\frac{1}{A(0)}\,\beta+{\cal O}(\beta^2)$ 
for $A(0)=U''_Q(0)$ and $e_{{\rm min}}(\beta)=U_Q(0)+{\cal O}(\beta^{1/2})$ 
as $\beta\to 0^+$. 
\end{exercise} 
{\bf Solution\,:} Due to \cite[Lemma A.7(f)]{MK} one has 
\[ r_0(\beta)^4=\frac{1}{A(0)}\,\beta+{\cal O}(\beta^2)
   \quad\mbox{and}\quad e_{{\rm min}}(\beta)=U_Q(0)+{\cal O}(\beta^{1/2}) \] 
as $\beta\to 0^+$. Since $U_Q(0)<e_0$ (the cut-off energy), 
the condition $e_{{\rm min}}(\beta)\le e_0$ 
from the characterization of $J$ in (\ref{Jdom}) is satisfied 
with strict inequality at $\beta=0$. 
It follows that $[0, \eps]\subset J$, if $\eps>0$ is sufficiently small. 
{\hfill$\Box$}\bigskip 

\begin{exercise} Prove that $J$ is bounded. 
\end{exercise} 
{\bf Solution\,:} First, if $\beta\in J$, then $r_0(\beta)\le r_Q$, 
where ${\rm supp}\,\rho_Q=[0, r_Q]$. Otherwise we would have $r_0(\beta)>r_Q$ 
for some $\beta\in J\setminus\{0\}$. 
Since $r_Q$ is characterized by $U_Q(r_Q)=e_0$, this gives $U_Q(r_0(\beta))>e_0$, 
and consequently $\beta\le 2r_0(\beta)^2\,(e_0-U_Q(r_0(\beta))\le 0$, 
which is a contradiction. 
Then $r_0(\beta)\le r_Q$ for $\beta\in J$ in turn leads to the boundedness of $J$, owing to 
\[ \beta\le 2r_0(\beta)^2\,(e_0-U_Q(r_0(\beta))\le 2r_Q^2\,(e_0-U_Q(r_0(\beta))
   \le 2r_Q^2\,(e_0-U_Q(0)) \] 
uniformly for $\beta\in J$.  
{\hfill$\Box$}\bigskip 

\begin{exercise} Prove that $\beta_\ast=\max J$ satisfies $e_{{\rm min}}(\beta_\ast)=e_0$. 
\end{exercise} 
{\bf Solution\,:} In fact, at $\beta_\ast$ we must have 
$2r_0(\beta_\ast)^2\,(e_0-U_Q(r_0(\beta_\ast))=\beta_\ast$. Thus  
\[ e_{{\rm min}}(\beta_\ast)=U_{{\rm eff}}(r_0(\beta_\ast), \beta_\ast)
   =U_Q(r_0(\beta_\ast))+\frac{\beta_\ast}{2r_0(\beta_\ast)^2}=e_0, \] 
which is the claim. {\hfill$\Box$}\bigskip 

To summarize, since the condition on $e$ is $e_0\ge e\ge e_{{\rm min}}(\beta)$, 
we have shown that 
\begin{equation}\label{DDabo} 
   D=\{(\beta, e): \beta\in [0, \beta_\ast], e\in [e_{{\rm min}}(\beta), e_0]\}
\end{equation} 
and $K=[0, 2\pi]\times D$ for the support $K$ of $Q$ in terms of $e$ and $\beta$, 
and the lower boundary curve 
$[0, \beta_\ast]\ni\beta\mapsto e_{{\rm min}}(\beta)$ strictly increases 
from $U_Q(0)$ to $e_0$. 

\begin{figure}[H]
   \centering
   \includegraphics[angle=0,width=0.5\linewidth]{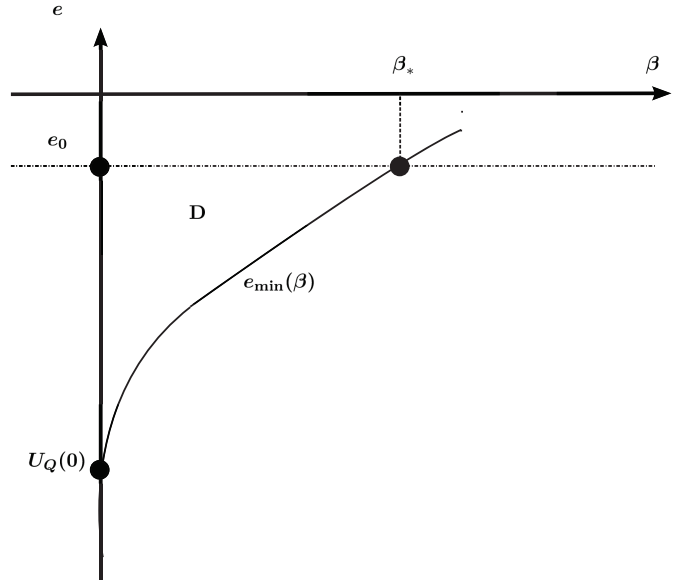}
   % \\[-10ex]
   \caption{The domain $D$ in coordinates $(e, \beta)=(E, \beta)$}
\end{figure}

\noindent
Observe that the reasoning in this section did not depend on the specific form 
of the polytropic ansatz function (\ref{poly_form}), but only on the general properties 
of the functions $r_0(\beta)$ and $e_{{\rm min}}(\beta)$. 
It should also be mentioned that $r_0(\beta_\ast)\in ]0, r_Q[$ is verified, 
see \cite[Section 1.7.1]{MK}. 
\medskip 

Going back to (\ref{zkm}) and using $\ell$ instead of $\beta$, we thus have 
\[ K=\{(\theta, E, \ell): \theta\in [0, 2\pi], \ell\in [0, l_\ast], 
   E\in [e_{{\rm min}}(\ell), e_0]\} \] 
in the variables $(\theta, E, \ell)$. Since $I=I(E, \ell)$ is the inverse function 
to $E=E(I, \ell)$ 
at fixed $\ell$, the set $K$ can be equally expressed in the variables $(\theta, I, \ell)$, 
which is the main observation here. As $\theta$ is $2\pi$-periodic, 
therefore spherically symmetric functions $g(x, v)=g(r, p_r, \ell)=g(\theta, I, \ell)$ 
of $(\theta, I, \ell)$ that are defined on $K$, the support of $Q$, 
can be expanded into a Fourier series 
\begin{equation}\label{bffdn} 
   g(\theta, I, \ell)=\sum_{k\in\Z} g_k(I, \ell)\,e^{ik\theta},
\end{equation} 
where $(I, \ell)\simeq (E, \ell)\simeq (E, \beta)\in D$. The Fourier coefficients are 
\[ g_k(I, \ell)=\frac{1}{2\pi}\int_0^{2\pi} g(\theta, I, \ell)\,e^{-ik\theta}\,d\theta. \] 

The series expansion (\ref{bffdn}) is most convenient, since one can easily 
do calculations on such series, or define Sobolev-type function space. 

This motivates the following 

\begin{definition}[$X^\alpha$-spaces] 
For $\alpha\ge 0$ denote 
\[ X^\alpha=\Big\{g=\sum_{k\in\Z} g_k(I, \ell)\,e^{ik\theta}: 
   {\|g\|}_{X^\alpha}^2=16\pi^3\sum_{k\in\Z} {(1+k^2)}^\alpha
   \,{\|g_k\|}^2_{L^2_{\frac{1}{|Q'|}}(D)}<\infty\Big\}, \] 
where 
\[ D=\{(E, \ell): \ell\in [0, \ell_\ast], E\in [e_{{\rm min}}(\ell), e_0]\} \]
is from (\ref{DDabo}) and expressed in $(I, \ell)$, and moreover 
\[ {(\phi, \psi)}_{L^2_{\frac{1}{|Q'|}}(D)}=\iint\limits_D dI\,d\ell\,\ell\,
   \frac{1}{|Q'(e)|}\,\overline{\phi(I, \ell)}\,\psi(I, \ell) \]  
is a weighted $L^2$-inner product for suitable functions $\phi, \psi$ on $D$; 
note $e=e(I, \ell)$. 
The associated scalar product on the Hilbert space $X^\alpha$ is given by
\[  {(g, h)}_{X^\alpha}
   =16\pi^3\sum_{k\in\Z} {(1+k^2)}^\alpha\,{(g_k, h_k)}_{L^2_{\frac{1}{|Q'|}}(D)} \] 
for $g=\sum_{k\in\Z} g_k\,e^{ik\theta}$ and $h=\sum_{k\in\Z} h_k\,e^{ik\theta}$.  
\end{definition} 

%%%%%%%%%%%%%%%%%%%%%%%%%%%%%%%%%%%%%%%%%%%%%%%%%%%%%%%%%%%%%%%%%%%%%%%%%%%%%%%

\section{Linearization} 
\setcounter{equation}{0}

Now that we have introduced some nice Hilbert spaces, we also need to have a suitable 
self-adjoint operator in order to come close to a possible Birman-Schwinger setting. 
Since we are interested in dynamical properties of the system close 
to an isotropic	 steady state 
solution $Q$, it is natural to consider the linearization about such a steady state. 

For, we write $f(t)=Q+g(t)$ with $g$ `small'. First note that 
\[ v\cdot\nabla_x f-\nabla_x U_f\cdot\nabla_v f=\{f, e_f\} \] 
for $e_f(x, v)=\frac{1}{2}\,|v|^2+U_f(x)$, where 
\[ \{g, h\}=\nabla_x g\cdot\nabla_v h-\nabla_v g\cdot\nabla_x h \] 
denotes the standard Poisson bracket of two functions $g=g(x, v)$ and $h=h(x, v)$. 

\begin{exercise}\label{poiscirc} 
Prove that if $\phi: \R\to\R$ is a function 
and $g=g(x, v)$, $h=h(x, v)$, then $\{\phi(g), h\}=\phi'(g)\{g, h\}$. 
\end{exercise} 

Therefore we may write the Vlasov equation (\ref{vpgr1}) as 
\begin{eqnarray*}
   0 & = & \partial_t f+\{f, e_f\}=\partial_t g+\Big\{Q+g, \frac{1}{2}\,|v|^2+U_Q+U_g\Big\}
   \\ & = & \partial_t g-\nabla_v Q\cdot\nabla_x U_g
   +v\cdot\nabla_x g-\nabla_v g\cdot\nabla_x U_Q-\nabla_v g\cdot\nabla_x U_g, 
\end{eqnarray*} 
which is equivalent to 
\begin{equation}\label{VPop} 
   \partial_t g+{\cal T}g+{\cal K}g=\nabla_v g\cdot\nabla_x U_g, 
\end{equation} 
where we have introduced the linear operators 
\begin{eqnarray} 
   {\cal T} g & = & v\cdot\nabla_x g-\nabla_v g\cdot\nabla_x U_Q=\{g, e_Q\}, 
   \label{taudef}
   \\ {\cal K} g & = & -\nabla_v Q\cdot\nabla_x U_g=\{Q, U_g\}; 
   \label{calKdef} 
\end{eqnarray} 
recall that $U_g=U_g(r)$, whence $\nabla_v U_g=0$. 
Since the term on the right-hand side of (\ref{VPop}) is (formally) quadratic in $g$, 
the linearization is found to be 
\begin{equation}\label{linVP_Q} 
   \partial_t g+{\cal T}g+{\cal K}g=0. 
\end{equation} 
\medskip 

The next step is to linearize not only the equation itself, 
but a suitable Lyapunov-type functional For this we will closely follow \cite{GRking} 
and and once again write $f(t)=Q+g(t)$. The total energy 
\[ {\cal H}(f(t))=\frac{1}{2}\int_{\R^3}\int_{\R^3} |v|^2\,f(t, x, v)\,dx\,dv
   -\frac{1}{8\pi}\int_{\R^3} |\nabla U_{f(t)}(t, x)|^2\,dx \] 
is conserved along solutions, so it could be suspected to be a Lyapunov function. 

\begin{exercise} Prove that $\frac{d}{dt}\,{\cal H}(f(t))=0$ for solutions $f(t)$ 
of the Vlasov-Poisson system. 
\end{exercise} 

The expansion about $Q$ then yields 
\begin{equation}\label{expa1} 
   {\cal H}(f(t))={\cal H}(Q)+\int_{\R^3}\int_{\R^3}
   e_Q\,g(t)\,dx\,dv
   -\frac{1}{8\pi}\int_{\R^3} |\nabla U_{g(t)}|^2\,dx+{\cal O}(g^3);
\end{equation}  
note that $f\mapsto U_f$ is linear. 

\begin{exercise} Prove that (\ref{expa1}) holds (formally). 
\end{exercise} 

The linear term on the right-hand side of (\ref{expa1}) does not vanish, 
i.e., $Q$ is not a critical point of ${\cal H}$. 
However, this defect can be remedied by making use of the fact that every 
`Casimir functional' 
\[ {\cal C}_\Phi(f(t))=\int_{\R^3}\int_{\R^3}\Phi(f(t, x, v))\,dx\,dv \] 
is also conserved along solutions, provided that $\Phi$ is sufficiently well-behaved. 
Passing from ${\cal H}$ to 
\[ {\cal H}_\Phi={\cal H}+{\cal C}_\Phi \] 
and repeating the expansion, one arrives at 
\begin{eqnarray}\label{tidy}  
   {\cal H}_\Phi(f(t)) 
   & = & {\cal H}_\Phi(Q)+\int_{\R^3}\int_{\R^3}(e_Q+\Phi'(Q))\,g(t)\,dx\,dv
   \nonumber
   \\ & & +\,\frac{1}{2}\int_{\R^3}\int_{\R^3}\Phi''(Q)\,g(t)^2\,dx\,dv
   -\frac{1}{8\pi}\int_{\R^3} |\nabla U_{g(t)}|^2\,dx+{\cal O}(g^3). 
\end{eqnarray} 
Writing $e=e_Q$, since $Q=Q(e)$, the equation $e+\Phi'(Q(e))=0$ can be (formally) solved 
by taking $\Phi'(\xi)=-Q^{-1}(\xi)$, at least if for instance $Q'(e)<0$ is verified 
for the relevant $e$ in the support of $Q$. 

\begin{exercise} For the polytropes (\ref{poly_form}), 
show that $\Phi(\xi)=\frac{k}{k+1}\,\xi^{\frac{k+1}{k}}-e_0\,\xi$, 
$\xi\in [0, \infty[$, is a possible choice. 
\end{exercise}

\noindent
Then $Q$ becomes a critical point of this ${\cal H}_\Phi$, 
and due to $1+\Phi''(Q(e))Q'(e)=0$ and $Q'(e)<0$ the expansion (\ref{tidy}) simplifies to 
\begin{eqnarray*} 
   {\cal H}_\Phi(f(t)) 
   & = & {\cal H}_\Phi(Q)+\frac{1}{2}\,{\cal A}(g(t), g(t))+{\cal O}(g^3),
   \\ {\cal A}(g, g) & = & \int_{\R^3}\int_{\R^3}\frac{dx\,dv}{|Q'(e_Q)|}\,|g|^2
   -\frac{1}{4\pi}\int_{\R^3} |\nabla_x U_g|^2\,dx. 
\end{eqnarray*} 
Thus one can expect that the stability of $Q$ will be determined by the properties 
of the quadratic (second variation) part ${\cal A}=2\,D^2 {\cal H}_\Phi(Q)$, 
which we will call the Antonov functional. 

\begin{exercise} Prove that $\frac{d}{dt}\,{\cal A}(g(t), g(t))=0$ along solutions 
$g(t)$ of the linearized equation (\ref{linVP_Q}). 
\end{exercise}

Ideally, to infer stability of $Q$ it would be helpful 
if ${\cal A}$ had some kind of coercivity property. 
Now it is the content of the celebrated Antonov stability estimate 
\cite{anton1,anton2}, that 
\begin{equation}\label{anton_prel} 
   {\cal A}({\cal T}u, {\cal T}u)\ge c {\|u\|}_Q^2
\end{equation} 
holds for all functions $u=u(x, v)$ that are spherically symmetric and odd in $v$, 
i.e., they satisfy $u(x, -v)=-u(x, v)$; the constant $c>0$ does only depends upon $Q$. 
The weighted $L^2$-inner product is defined as 
\begin{equation}\label{innerQ} 
   {(g, h)}_Q=\iint\limits_K\frac{1}{|Q'(e_Q)|}\,\overline{g(x, v)}\,h(x, v)\,dx\,dv 
\end{equation}  
and it induces the norm ${\|\cdot\|}_Q$. 
Perturbations of the form $g={\cal T}u$ are called `dynamically accessible', 
for reasons explained in \cite{morrison}, also see \cite{Perez}.
Antonov \cite{anton1,anton2} could prove that the positive definiteness 
(\ref{anton_prel}) is equivalent to the linear stability of $Q$.  
Many works followed these pioneering observations, and until to date almost all 
stability proofs, linear or nonlinear, use the Antonov stability estimate 
in one way or another. The bound (\ref{anton_prel}), or variations thereof, 
is applied in a number of papers, both in the physics and in the mathematics community, 
to address a variety of stability issues; 
see \cite{DBF,GCDB,GuoAnt,GRking,Kand,KS,LMR11,LMR12,MP,SdLRP} 
and many further. 
\medskip 

In view of (\ref{anton_prel}), we first need to obtain a better understanding 
of spherically symmetric functions $g$ that are odd in $v$. 

\begin{exercise} Prove the following facts: 
\begin{itemize} 
\item[(a)] If $(x, v)\mapsto (p_r, L_3, \ell; r, \varphi, \chi)$ 
under the above transformation (\ref{1stchange}), then we have $(x, -v)
\mapsto (-p_r, -L_3, \ell; r, \varphi+\pi, \pi-\chi)$. 
\item[(b)] If 
\[ (x, v)\mapsto (p_r, L_3, \ell; r, \varphi, \chi)
   \mapsto (I, L_3, \ell; \theta, \varphi, \chi) \] 
under the transformation (\ref{oacan}), then 
\[ (x, -v)\mapsto (-p_r, -L_3, \ell; r, \varphi+\pi, \pi-\chi)
   \mapsto (I, -L_3, \ell; 2\pi-\theta, \varphi+\pi, \pi-\chi). \] 
\item[(c)] $g$ is even in $v$ if and only if $g(r, -p_r, \ell)=g(r, p_r, \ell)$ 
if and only if $g(2\pi-\theta, I, \ell)=g(\theta, I, \ell)$. 
\item[(d)] $g$ is odd in $v$ if and only if $g(r, -p_r, \ell)=-g(r, p_r, \ell)$ 
if and only if $g(2\pi-\theta, I, \ell)=-g(\theta, I, \ell)$. 
\item[(e)] $g(\theta, I, \ell)=\sum_{k\in\Z} g_k(I, \ell)\,e^{ik\theta}$ 
is even in $v$ if and only if $g_{-k}=g_k$ for all $k\in\Z$. 
If $g$ is real-valued, then $g_k(I, \ell)\in\R$. 
\item[(f)] $g(\theta, I, \ell)=\sum_{k\in\Z} g_k(I, \ell)\,e^{ik\theta}$ 
is odd in $v$ if and only if $g_{-k}=-g_k$ for all $k\in\Z$, 
and in particular $g_0=0$. If $g$ is real-valued, then $g_k(I, \ell)\in i\R$. 
\end{itemize}
\end{exercise}

\noindent 
Also note that necessarily $g_0=0$ for odd functions.  

\begin{definition}[$X^\alpha_{{\rm odd}}$-spaces]
For $\alpha\ge 0$ denote 
\[ X^\alpha_{{\rm odd}}=\{g\in X^\alpha: g_{-k}=-g_k\,\,for\,\,k\in\Z\}. \] 
\end{definition} 

Now we are in a position to introduce one of our main objects of interest, 
namely the operator 
\begin{equation}\label{Ldef}
   Lu = -{\cal T}^2 u-{\cal K}{\cal T}u. 
\end{equation} 
The connection to the stability problem outlined above is made in 

\begin{lemma}\label{L11}  
$L$ is self-adjoint on the domain ${\cal D}(L)=X^2_{{\rm odd}}$ in $X^0_{{\rm odd}}$. 
In addition, 
\[ {(Lu, u)}_Q={\cal A}({\cal T}u, {\cal T}u) \] 
holds for $u\in X^2_{{\rm odd}}$. 
\end{lemma}  

\noindent 
See \cite[Lemma 1.1]{MK} for the proof. Here we only give a somewhat rough argument, why 
\begin{equation}\label{kermi} 
   {(-{\cal T}^2 u, u)}_Q
   =\int_{\R^3}\int_{\R^3}\frac{dx\,dv}{|Q'(e_Q)|}\,|{\cal T}u|^2
\end{equation} 
can be expected to hold; think of ${(-\Delta u, u)}_{L^2}={\|\nabla u\|}^2_{L^2}$ 
under appropriate hypotheses on $u$. First we observe that ${\cal T}$ from (\ref{taudef}) 
can be written as 
\[ {\cal T} g=v\cdot\nabla_x g-\nabla_v g\cdot\nabla_x U_Q
   ={\rm div}_x(vg)-{\rm div}_v(g\nabla_x U), \] 
since $U_Q$ is independent of $v$. Therefore 
\[ \int_{\R^3}\int_{\R^3} ({\cal T}g)h\,dx\,dv=-\int_{\R^3}\int_{\R^3} 
   g({\cal T}h)\,dx\,dv \] 
through integration by parts, if there are no boundary terms. 
Thus if $u$ is real-valued and has its support in $K$, then we have by (\ref{innerQ})
\[ {(-{\cal T}^2 u, u)}_Q=-\iint\limits_K\frac{1}{|Q'(e_Q)|}\,({\cal T}^2 u)\,u\,dx\,dv
   =\int_{\R^3}\int_{\R^3} ({\cal T} u)\,{\cal T}\Big(\frac{1}{|Q'(e_Q)|}\,u\Big)\,dx\,dv. \] 
Clearly ${\cal T}$ satisfies the product rule ${\cal T}(gh)=({\cal T}g)h+g({\cal T}h)$. 
Now if we use (\ref{taudef}) and Exercise \ref{poiscirc} for $\phi(s)=\frac{1}{|Q'(s)|}$, 
then we obtain 
\[ {\cal T}\Big(\frac{1}{|Q'(e_Q)|}\Big)=\{\phi(e_Q), e_Q\}=\phi'(e_Q)\{e_Q, e_Q\}=0, \] 
and (\ref{kermi}) follows. 
\medskip 

Due to the Antonov bound (\ref{anton_prel}) and Lemma \ref{L11}, 
the spectrum of the second variation (in dynamically accessible directions) 
is expected to play a major role in all kinds of stability questions. 

%%%%%%%%%%%%%%%%%%%%%%%%%%%%%%%%%%%%%%%%%%%%%%%%%%%%%%%%%%%%%%%%%%%%%%%%%%%%%%%

\section{The Birman-Schwinger approach} 
\setcounter{equation}{0}

Therefore the task is to extract as much information on the spectrum of $L$ as possible.  
To begin with, we recall that the discrete spectrum 
of a self-adjoint operator $L$ in a Hilbert space, called $\sigma_d(L)$, 
consists of all eigenvalues of $L$ of finite multiplicity that are isolated points 
of the spectrum $\sigma(L)$. Its complement $\sigma_{{\rm ess}}(A)
=\sigma(A)\setminus\sigma_d(L)$ 
is the essential spectrum. 

Let us first consider this part of the spectrum for 
$Lu=-{\cal T}^2 u-{\cal K}{\cal T}u$ from (\ref{Ldef}) 
on ${\cal D}(L)=X^2_{{\rm odd}}$ in the Hilbert space $X^0_{{\rm odd}}$. 

\begin{lemma} The following assertions hold: 
\begin{itemize} 
\item[(a)] $-{\cal T}^2: X^2_{{\rm odd}}\to X^0_{{\rm odd}}$ is a self-adjoint operator. 
\item[(b)] The operator ${\cal K}{\cal T}: X^0_{{\rm odd}}\to X^0_{{\rm odd}}$ 
is given by 
\[ {\cal K}{\cal T}g=4\pi\,|Q'(e_Q)|\,p_r\int_{\R^3} p_r\,g\,dv \] 
and it is linear, bounded, 
symmetric and positive: 
\begin{equation}\label{KTprod} 
   {({\cal K}{\cal T}g, g)}_{X^0}=\frac{1}{4\pi}\int_{\R^3} |U'_{{\cal T}g}(r)|^2\,dx\ge 0
   \quad\mbox{for}\quad g\in X^0.
\end{equation} 
\item[(c)] ${\cal K}{\cal T}$ is relatively $L$-compact, 
in that ${\cal D}({\cal K}{\cal T})=X^0\supset X^2_{{\rm odd}}={\cal D}(L)$ for the domains 
and ${\cal K}{\cal T}(L+i)^{-1}: X^0_{{\rm odd}}\to X^0_{{\rm odd}}$ 
is a compact operator.
\item[(d)] We have 
\[ \sigma_{{\rm ess}}(L)=\sigma_{{\rm ess}}(-{\cal T}^2). \] 
\end{itemize} 
\end{lemma} 
{\bf Proof\,:} See \cite[Cor.~B.10]{MK} for (a), \cite[Cor.~B.15]{MK} for (b) 
and the proof of \cite[Cor.~B.19]{MK} for (c). 
Essentially this is due to the fact that ${\cal K}: X^0\to X^0$ is compact, 
\cite[Cor.~C.6]{MK}. To establish the latter property, one uses that 
${\cal K}g=|Q'(e_Q)| p_r U'_g(r)$ 
(see below) and 
\[ {\|\nabla (\nabla U_g)\|}_{L^2(\R^3)}\le C{\|
   \nabla (\nabla\Delta^{-1}\rho_g)\|}_{L^2(\R^3)}
   \le C{\|\rho_g\|}_{L^2(\R^3)} \] 
together with the Sobolev embedding theorem; the regularizing property 
of $\Delta U_g=4\pi\rho_g$ 
is central to many stability results for Vlasov-Poisson. 
(d) This is a consequence of Weyl's Theorem, 
cf.~\cite[Thm.~14.6]{HiSi}. 
{\hfill$\Box$}\bigskip

Thus we need to determine the essential spectrum of $-{\cal T}^2$.
For this, the variables $(\theta, I, \ell)$ are most convenient, 
and we are going to use the fact that canonical transformations 
leave Poisson brackets unaltered. Hence if we write 
$\Phi=(r, \varphi, \chi)$, $\mathfrak{A}=(p_r, L_3, \ell)$, 
$\Theta=(\theta, \varphi, \chi)$ and $\mathfrak{I}=(I, L_3, \ell)$ for the coordinates, 
see (\ref{oacan}), then 
\[ {\{g, h\}}_{xv}={\{g, h\}}_{\Phi\mathfrak{A}}={\{g, h\}}_{\Theta\mathfrak{I}}. \] 
But the functions do depend only upon $(r, p_r, \ell)$ 
and $(\theta, I, \ell)$, respectively. Thus 
\begin{eqnarray*} 
   {\{g, h\}}_{\Phi\mathfrak{A}} & = & (\partial_r g)(\partial_{p_r} h)
   -(\partial_r h)(\partial_{p_r} g),
   \\ {\{g^\ast, h^\ast\}}_{\Theta\mathfrak{I}} & = & (\partial_\theta g)(\partial_I h)
   -(\partial_\theta h)(\partial_I g).  
\end{eqnarray*}    
Next we recall that $e_Q=e=E(I, \ell)$. 
Hence due to $\omega_1=\frac{\partial E}{\partial I}$ and $\partial_\theta E=0$ we get 
\[ {\cal T}g=\{g, e_Q\}=(\partial_\theta g)(\partial_I E)-(\partial_\theta E)(\partial_I g)
   =\omega_1\partial_\theta g, \] 
which is appealingly simple in the coordinates $(\theta, I, \ell)$. 
Since $\omega_1$ is independent of $\theta$, see (\ref{om123}), it also follows that 
\[ -{\cal T}^2 g =-\omega_1\partial_\theta (\omega_1\partial_\theta g)
   =-\omega_1^2\,\partial^2_\theta g. \] 
This relation makes it clear that the properties of the function 
$\omega_1=\omega_1(e, \ell)=\omega_1(e, \beta)$, 
or equivalently of the period function $T_1=\frac{2\pi}{\omega_1}$, on $D$ will be important. 

\begin{lemma} The following assertions hold: 
\begin{itemize} 
\item[(a)] We have $\omega_1\in C^1(\mathring{D})\cap C(D)$.  
\item[(b)] It holds that 
\[ \inf\omega_1=\delta_1>0\quad\mbox{and}\quad\sup\omega_1=\Delta_1<\infty. \] 
\end{itemize}  
\end{lemma} 
{\bf Proof\,:} See \cite[Thm.~3.6\,\& Thm.~3.13]{MK} for (a) 
and \cite[Thm.~3.2\,\& Thm.~3.5]{MK} for (b). 
Since $\omega_1$ is continuous on the compact set $D$ and $T_1$ is non-zero, 
certainly (b) follows from (a), but it is also possible to give a proof 
using the explicit period relation (\ref{T1form}). 
Similar results have been obtained in \cite{BAH}. 
{\hfill$\Box$}\bigskip

There is a result in the spectral theory of self-adjoint operators 
that asserts that the spectrum of the multiplication operator 
$M: D(M)=\{u\in L^2(\R^n): \chi u\in L^2(\R^n)\}\to L^2(\R^n)$, 
$Mu=\chi u$, for a given real-valued and continuous function $\chi$ 
has spectrum the $\overline{{\rm ran}\,\chi}$. 
If we also take into account that on a function $g=\sum_{k\in\Z} g_k\,e^{ik\theta}$ 
with coefficient $g_k=g_k(I, \ell)$ we have 
\[ -{\cal T}^2: g\cong (g_k)\mapsto (\omega_1^2\,k^2 g_k) \] 
then the following characterization of the essential spectrum of $-{\cal T}^2$, 
and thus of the one of $L$, is not a big surprise. 

\begin{theorem}\label{zordo} For the essential spectrum we have 
\[ \sigma_{{\rm ess}}(L)=\sigma_{{\rm ess}}(-{\cal T}^2)
   =\bigcup_{k=1}^\infty k^2 [\delta_1^2, \Delta_1^2]
   \quad\mbox{and}\quad\delta_1^2=\min\sigma_{{\rm ess}}(L)>0. \] 
If $\omega_1$ is not constant, then there exists $\lambda_c>\delta_1^2$ 
such that $[\lambda_c, \infty[\subset\sigma_{{\rm ess}}(L)$.  
\end{theorem} 
{\bf Proof\,:} See \cite[Cor.~B.19]{MK}. 
{\hfill$\Box$}\bigskip
 
\begin{exercise} Prove the last statement of Theorem \ref{zordo}. 
\end{exercise}

What we have done so far is more or less standard, 
but now we are getting closer to the heart of the matter. 
As before, we are trying to understand the spectrum of $L$, 
but this time, more specifically, possible eigenvalues below the essential spectrum; 
the following calculation is motivated by \cite{mathur}. 
Let $\lambda<\delta_1^2$ and suppose that $Lu=\lambda u$ for some 
$u\in X^2_{{\rm odd}}$ and $u\neq 0$. 
Then $(-{\cal T}^2-\lambda)u={\cal K}{\cal T}u$. 
Defining $\psi=(-{\cal T}^2-\lambda)u\in X^0_{{\rm odd}}$, we get 
\begin{equation}\label{stev} 
   \psi={\cal K}{\cal T}(-{\cal T}^2-\lambda)^{-1}\psi.
\end{equation} 
Now 
\[ {\cal K} g=-\nabla_v Q\cdot\nabla_x U_g
   =-Q'(e_Q)v\cdot U'_g(r)\frac{x}{r}=|Q'(e_Q)|\,p_r\,U'_g(r) \] 
by the definition of ${\cal K}$ in (\ref{calKdef}) and since $Q'(e_Q)<0$. 
Thus the image of the operator ${\cal K}$ is special. 
Apart from the factor $|Q'(e_Q)|$, it consists of function $h=h(r, p_r, \ell)$ 
that factorize as $p_r\tilde{h}(r)$. In particular, 
due to (\ref{stev}), also $\psi$ can be written in this way and we obtain 
\begin{equation}\label{eul} 
   \psi={\cal K}{\cal T}u=|Q'(e_Q)|\,p_r\,U'_{{\cal T}u}(r).
\end{equation} 

To make sense of the following definition, we need to mention that 
in general for spherically symmetric functions $g$ one has 
$U'_{{\cal T}g}(r)=4\pi\int_{\R^3} p_r\,g\,dv$. 

\begin{exercise} Prove this, using that $\rho_{{\cal T}g}(x)={\rm div}_x\int_{\R^3} v\,g\,dv$ 
and Gauss's Theorem in $U'_h(r)=\frac{1}{r^2}\int_{|x|\le r} \rho_h(x)\,dx$. 
\end{exercise}

\begin{definition}[The Birman-Schwinger operators] 
Let 
\begin{equation}\label{wvh} 
   {\cal Q}_\lambda\Psi=U'_{{\cal T}(-{\cal T}^2-\lambda)^{-1}\psi}
   =4\pi\int_{\R^3} p_r\,(-{\cal T}^2-\lambda)^{-1}\psi\,dv
\end{equation} 
for functions $\Psi=\Psi(r)$, where we put $\psi(r, p_r, \ell)=|Q'(e_Q)|\,p_r\,\Psi(r)$ 
in terms of a given $\Psi$. 
\end{definition} 

\noindent 
Since $\int dv$ is integrated out in (\ref{wvh}), it turns out 
that ${\cal Q}_\lambda\Psi=({\cal Q}_\lambda\Psi)(r)$ is also a function of $r$ only.  
Coming back to the spectral problem for $L$, we started out with an eigenfunction $u$ of $L$, 
thereafter put $\psi=(-{\cal T}^2-\lambda)u$, and now let 
\[ \Psi(r)=U'_{{\cal T}(-{\cal T}^2-\lambda)^{-1}\psi}(r)=U'_{{\cal T}u}(r) \] 
to deduce 
\[ |Q'(e_Q)|\,p_r\,\Psi(r)=|Q'(e_Q)|\,p_r\,U'_{{\cal T}u}(r)=\psi \] 
from (\ref{eul}). Therefore we obtain 
\[ {\cal Q}_\lambda\Psi=U'_{{\cal T}(-{\cal T}^2-\lambda)^{-1}\psi}=\Psi. \] 
In other words, $1$ is an eigenvalue of ${\cal Q}_\lambda$ 
with eigenfunction $\Psi$. Since a converse statement can be verified 
similarly, we arrive at 

\begin{theorem}\label{EWchar} Let $\lambda<\delta_1^2$. 
Then $\lambda$ is an eigenvalue of $L$ if and only if $1$ 
is an eigenvalue of ${\cal Q}_\lambda$. More precisely, 
\begin{itemize}
\item[(a)] if $u\in X^2_{{\rm odd}}$ is an eigenfunction of $L$ 
for the eigenvalue $\lambda$, then $\Psi=U'_{{\cal T}u}\in L^2_r$ 
is an eigenfunction of ${\cal Q}_\lambda$ for the eigenvalue $1$; 
\item[(b)] if $\Psi\in L^2_r$ is an eigenfunction of ${\cal Q}_\lambda$ 
for the eigenvalue $1$, 
then $u=(-{\cal T}^2-\lambda)^{-1}(|Q'(e_Q)|\,p_r\Psi)\in X^2_{{\rm odd}}$ 
is an eigenfunction of $L$ for the eigenvalue $\lambda$. 
\end{itemize} 
\end{theorem} 
{\bf Proof\,:} See \cite[Thm.~4.5]{MK}. 
{\hfill$\Box$}\bigskip

Here $L^2_r$ denotes the $L^2$-Lebesgue space of radially 
symmetric functions $\Psi(x)=\Psi(r)$ 
on $\R^3$, where we take 
\[ \langle\Psi, \Phi\rangle=\int_{\R^3}\overline{\Psi(x)}\,\Phi(x)\,dx
   =4\pi\int_0^\infty r^2\,\overline{\Psi(r)}\,\Phi(r)\,dr \] 
as the inner product of $\Psi, \Phi\in L^2_r$.

\begin{exercise} Prove part (b) of Theorem \ref{EWchar}. 
\end{exercise}

If we compare Theorem \ref{EWchar} to the quantum mechanics result 
Theorem \ref{BSinQM}, then we see that also in galactic dynamics 
there is a Birman-Schwinger principle. 
Furthermore, it is nice to observe that both are formally identical, 
if we associate $p_r\sim\sqrt{-V}$ and $-\Delta\sim -{\cal T}^2$, 
and furthermore  disregard the velocity average $\int_{\R^3} dv$; 
the appearance of $|Q'(e_Q)|$ in $|Q'(e_Q)|\,p_r\Psi$ is due to the 
${(\cdot, \cdot)}_Q$ that is used. There is yet another fact 
that supports the analogy of both approaches. The operator ${\cal Q}_\lambda$ 
from (\ref{wvh}) can be expressed as 
\[ {\cal Q}_\lambda\Psi
   =4\pi\int_{\R^3} p_r\,(-{\cal T}^2-\lambda)^{-1}\,(|Q'(e_Q)|\,p_r\Psi)\,dv. \] 
Comparing this relation to (\ref{Be}), it turns out that both relations do agree, 
if we apply the same identifications as before. 
\medskip 

Theorem \ref{EWchar} could only be useful if we are able to gain a better understanding 
of the operators ${\cal Q}_\lambda$, which turn out to have a couple 
of nice properties. One also notices that ${\cal Q}_z$ can not only be defined 
for $z=\lambda<\delta_1^2$, 
but for all $z\in\Omega=\C\setminus [\delta_1^2, \infty[$. 
 
\begin{lemma}[Properties of ${\cal Q}_z$]
The following assertions hold. 
\begin{itemize}
\item[(a)] For every $z\in\Omega$ we have ${\cal Q}_z\in {\cal B}(L^2_r)$, 
the space of linear and bounded operators on $L^2_r$. In addition, the map 
\[ \Omega\ni z\mapsto {\cal Q}_z\in {\cal B}(L^2_r) \] 
is analytic, and we have the representation 
\begin{eqnarray}\label{2Q}  
    ({\cal Q}_z\Psi)(r) & = & \frac{16\pi}{r^2}\sum_{k\neq 0}\int_0^\infty 
    d\tilde{r}\,\Psi(\tilde{r})
   \iint\limits_D d\ell\,\ell\,de\,{\bf 1}_{\{r_-(e,\,\ell)\le r,\,\tilde{r}
   \le r_+(e,\,\ell)\}}
   \,\frac{\omega_1(e, \ell)\,|Q'(e)|}{(k^2\omega_1^2(e, \ell)-z)} 
   \nonumber
   \\ & & \hspace{14em}\times\sin(k\theta(r, e, \ell))\sin(k\theta(\tilde{r}, e, \ell))
\end{eqnarray} 
for $\Psi\in L^2_r$. 
\item[(b)] If $z\in\Omega$, then 
\[ ({\cal Q}_z\Psi)(r)=\langle K_{\bar{z}}(r, \cdot), \Psi\rangle \]
for $\Psi\in L^2_r$. The integral kernel $K_z$ is given by 
\begin{eqnarray*}  
   \lefteqn{K_z(r, \tilde{r})}
   \\ & = & \frac{4}{r^2\tilde{r}^2}\sum_{k\neq 0}
   \iint\limits_D d\ell\,\ell\,de\,{\bf 1}_{\{r_-(e,\,\ell)\le r,\,\tilde{r}
   \le r_+(e,\,\ell)\}}
   \,\frac{\omega_1(e, \ell)\,|Q'(e)|}{k^2\omega_1^2(e, \ell)-z}
   \,\sin(k\theta(r, e, \ell))\sin(k\theta(\tilde{r}, e, \ell)).
   \\ 
\end{eqnarray*} 
\item[(c)] If $z\in\Omega$, then ${\cal Q}_z$ is a Hilbert-Schmidt operator on $L^2_r$.  
\item[(d)] If $\lambda\in ]-\infty, \delta_1^2[$, then ${\cal Q}_\lambda$ is symmetric 
and positive. Its spectrum consists of $\mu_1(\lambda)\ge\mu_2(\lambda)\ge\ldots\to 0$ 
(the eigenvalues are listed according to their multiplicities). In addition, 
\[ \mu_1(\lambda)=\|{\cal Q}_\lambda\|
   =\sup\,\{\langle {\cal Q}_\lambda\Psi, \Psi\rangle: {\|\Psi\|}_{L^2_r}\le 1\} \] 
for the largest eigenvalue of ${\cal Q}_\lambda$, 
where $\|\cdot\|={\|\cdot\|}_{{\cal B}(L^2_r)}$. 
\end{itemize} 
\end{lemma}  
{\bf Proof\,:} See \cite[Lemma 4.3]{MK}. The representation formula 
(\ref{2Q}) is very convenient and obtained from (\ref{wvh}) 
by Fourier expanding the functions involved and using the fact that 
$\psi(r, p_r, \ell)=|Q'(e_Q)|\,p_r\,\Psi(r)$ has 
\begin{equation}\label{psicoeff} 
   \psi_k(I, \ell) = -\frac{i}{\pi}\,|Q'(e)|\,\omega_1(e, \ell)
   \int_{r_-(e, \ell)}^{r_+(e, \ell)} d\tilde{r}\,\Psi(\tilde{r})
   \sin(k\theta(\tilde{r}, e, \ell))
\end{equation} 
as its Fourier coefficients. 
{\hfill$\Box$}\bigskip 

\begin{exercise} Prove (\ref{psicoeff}) from (\ref{rprsymm}). 
\end{exercise}

According to Theorem \ref{EWchar}, in order to find eigenvalues 
$\hat{\lambda}<\delta_1^2$, we have to locate such a $\hat{\lambda}$ 
that additionally satisfies $\mu_1(\hat{\lambda})=1$. 
Therefore we have to study the function 
$\mu_1: ]-\infty, \delta_1^2[\to ]0, \infty[$ 
in more detail. 

\begin{lemma} We have $0<\mu_1(0)<1$, and $\mu$ 
is monotone increasing, convex and locally Lipschitz continuous. 
The limit 
\begin{equation}\label{iastdef} 
   \mu_\ast=\lim_{\lambda\to\delta_1^2-}\mu_1(\lambda)
   =\sup\,\{\mu_1(\lambda): \lambda\in [0, \delta_1^2[\} 
   \in [\mu_1(0), \infty]
\end{equation}  
does exist. 
\end{lemma} 
{\bf Proof\,:} See \cite[Lemma 4.3\,\& Lemma 4.7(a), (d)]{MK}. 
{\hfill$\Box$}\bigskip 

\section{An application} 
\setcounter{equation}{0}

It is to be expected that a good understanding of the 
Birman-Schwinger operators ${\cal Q}_z$ and their spectra 
will lead to new insights into stability-related properties of solutions 
close to a static solution of the Vlasov-Poisson system. 
\medskip 

As an example application, we consider 
\[  \lambda_\ast=\inf\,\{{(Lu, u)}_Q: u\in X^2_{{\rm odd}}, {\|u\|}_Q=1\}>0, \] 
which is the `best constant' in the Antonov stability estimate (\ref{anton_prel}); 
recall Lemma \ref{L11}. In \cite{MK} we derived many results 
related to $\lambda_\ast$, and in particular we we able to characterize 
the cases where $\lambda_\ast$ is attained, 
in the sense that $\lambda_\ast={(Lu_\ast, u_\ast)}_Q$ for some minimizing function 
$u_\ast\in X^2_{{\rm odd}}$ 
such that ${\|u_\ast\|}_Q=1$. It turns out that then $u_\ast$ will be an eigenfunction of $L$ 
corresponding to the eigenvalue $\lambda_\ast$, so that $Lu_\ast=\lambda_\ast u_\ast$. 
Both $u_\ast$ and the quantity $\lambda_\ast$ will be of fundamental importance 
for the dynamics of the gravitational Vlasov-Poisson system. 

\begin{lemma} 
Let $u_\ast\in X^2_{{\rm odd}}$ be a minimizer and define 
\[ g_\ast(t, x, v)=\cos(\sqrt{\lambda_\ast}t)\,u_\ast(x, v)
   -\frac{1}{\sqrt{\lambda_\ast}}\,\sin(\sqrt{\lambda_\ast}t)\,({\cal T}u_\ast)(x, v). \]  
Then $g_\ast$ is a $\frac{2\pi}{\sqrt{\lambda_\ast}}$-periodic solution of the equation 
(\ref{linVP_Q}) that is obtained by linearizing Vlasov-Poisson about $Q$. 
\end{lemma} 
{\bf Proof\,:} Observe that $u_\ast$ is odd in $v$. 
Hence $\rho_{u_\ast}(x)=\int_{\R^3} u_\ast(x, v)\,dv=0$ implies that 
$U_{u_\ast}=4\pi\Delta^{-1}\rho_{u_\ast}=0$ 
and therefore ${\cal K}u_\ast=0$ by (\ref{calKdef}). Consequently, 
\begin{eqnarray*} 
   \lefteqn{\partial_t g_\ast+{\cal T}g_\ast+{\cal K}g_\ast} 
   \\ & = & -\sqrt{\lambda_\ast}\sin(\sqrt{\lambda_\ast}t)\,u_\ast
   -\cos(\sqrt{\lambda_\ast}t)\,{\cal T}u_\ast
   +\cos(\sqrt{\lambda_\ast}t)\,{\cal T}u_\ast
   -\frac{1}{\sqrt{\lambda_\ast}}\,\sin(\sqrt{\lambda_\ast}t)\,{\cal T}^2 u_\ast
   \\ & & +\,\cos(\sqrt{\lambda_\ast}t)\,{\cal K}u_\ast
   -\frac{1}{\sqrt{\lambda_\ast}}\,\sin(\sqrt{\lambda_\ast}t)\,{\cal K}{\cal T}u_\ast
   \\ & = & -\sqrt{\lambda_\ast}\sin(\sqrt{\lambda_\ast}t)\,u_\ast
   +\frac{1}{\sqrt{\lambda_\ast}}\,\sin(\sqrt{\lambda_\ast}t)\,Lu_\ast
   \\ & = & 0, 
\end{eqnarray*} 
as claimed. 
{\hfill$\Box$}\bigskip 

Next we will clarify where $\lambda_\ast$ 
is located as compared to $\delta_1^2$, 
which is the minimum of the essential spectrum of $L$; 
recall Theorem \ref{zordo}. 

\begin{lemma} 
We have $\lambda_\ast\le\delta_1^2$. 
\end{lemma} 
{\bf Proof\,:} See \cite[Lemma 3.18]{MK}. The result is at least conceivable 
from the following observation: since $Lu=-{\cal T}^2 u-{\cal K}{\cal T}u$, 
using (\ref{kermi}) and (\ref{KTprod}) we get 
\begin{eqnarray*} 
   {(Lu, u)}_Q & = & {(-{\cal T}^2 u, u)}_Q-{({\cal K}{\cal T}u, u)}_Q 
   \\ & = & \int_{\R^3}\int_{\R^3}\frac{dx\,dv}{|Q'(e_Q)|}\,|{\cal T}u|^2
   -\frac{1}{4\pi}\int_{\R^3} |\nabla_x U_{{\cal T}u}|^2\,dx
   \\ & \le &  \int_{\R^3}\int_{\R^3}\frac{dx\,dv}{|Q'(e_Q)|}\,|{\cal T}u|^2. 
\end{eqnarray*} 
The latter expression equals ${(-{\cal T}^2 u, u)}_Q$, and it is just 
the quadratic form associated to $-{\cal T}^2$. 
One can construct suitable $u_j\in X^2_{{\rm odd}}$ such that ${\|u_j\|}_Q=1$ 
and ${(-{\cal T}^2 u_j, u_j)}_Q\to\delta_1^2$ as $j\to\infty$. 
{\hfill$\Box$}\bigskip
 
For the remaining part of these lectures, we will deal with the following result 
that illustrates the usefulness of the Birman-Schwinger operators. 

\begin{theorem}\label{iastgr} We have 
\[ \mu_\ast>1\,\,\Longleftrightarrow\,\,\lambda_\ast<\delta_1^2. \] 
In this case $\mu_1(\lambda_\ast)=1$ and $\lambda_\ast$ is an eigenvalue of $L$. 
\end{theorem} 
{\bf Proof\,:} See \cite[Thm.~4.13]{MK}. 
{\hfill$\Box$}\bigskip
  
It is not too hard to show that if $\mu_\ast>1$, then $\lambda_\ast=\delta_1^2$ 
is impossible, 
so that we must have $\lambda_\ast<\delta_1^2$. The converse statement 
is much more difficult to prove. 
Thus let us suppose that $\lambda_\ast<\delta_1^2$ holds, and assume that we already knew 
that $\lambda_\ast$ were an eigenvalue of $L$. Let $u_\ast\in X^2_{{\rm odd}}$ 
denote an associated eigenfunction. 
Using Theorem \ref{EWchar}(a), it follows that $\Psi_\ast=U'_{{\cal T}u_\ast}\in L^2_r$ 
is an eigenfunction of ${\cal Q}_{\lambda_\ast}$ for the eigenvalue $1$. 
Since $\mu_1(\lambda_\ast)$ 
is the largest eigenvalue of ${\cal Q}_{\lambda_\ast}$, we get $\mu_1(\lambda_\ast)\ge 1$. 
From the Antonov stability estimate ${(Lg, g)}_Q\ge\lambda_\ast{\|g\|}_Q^2$ 
it can be moreover deduced that $\mu_1(\lambda_\ast)\le 1$ is verified; 
see \cite[Lemma 4.7(b)]{MK}. 
Hence we obtain $\mu_1(\lambda_\ast)=1$ 
and it remains to show that $\mu_\ast>1$. Suppose that on the contrary $\mu_\ast\le 1$ 
is satisfied. 
For $\lambda\in [\lambda_\ast, \delta_1^2[$ the monotonicity of $\mu_1$ then yields 
$1=\mu_1(\lambda_\ast)\le\mu_1(\lambda)\le\mu_\ast\le 1$, which means that $\mu_1(\lambda)=1$ 
is constant for $\lambda\in [\lambda_\ast, \delta_1^2[$. Fixing normalized eigenfunctions 
$\Psi_{\tilde{\lambda}}$ for $\mu_1(\tilde{\lambda})$, 
where $\lambda_\ast\le\tilde{\lambda}<\lambda<\delta_1^2$, 
we find 
\[ 1=\mu_1(\tilde{\lambda})=\langle {\cal Q}_{\tilde{\lambda}}
   \Psi_{\tilde{\lambda}}, \Psi_{\tilde{\lambda}}\rangle
   \le\langle {\cal Q}_\lambda\Psi_{\tilde{\lambda}}, \Psi_{\tilde{\lambda}}\rangle
   \le\|{\cal Q}_\lambda\|\,{\|\Psi_{\tilde{\lambda}}\|}^2_{L^2_r}=\mu_1(\lambda)=1 \]  
from the general monotonicity of $\lambda\mapsto\langle 
{\cal Q}_\lambda\Psi, \Psi\rangle$, and therefore    
\[ \langle {\cal Q}_\lambda\Psi_{\tilde{\lambda}}, \Psi_{\tilde{\lambda}}\rangle=1, 
   \quad\lambda_\ast\le\tilde{\lambda}<\lambda<\delta_1^2. \] 
This can be shown to lead to a contradiction upon differentiation w.r.~to $\lambda$. 
\medskip 

To summarize the preceding argument, to establish ``$\Longleftarrow$'' 
in Theorem \ref{iastgr}, 
we need to prove that $\lambda_\ast<\delta_1^2$ implies that $\lambda_\ast$ 
is an eigenvalue of $L$. It turned out that this can be done by considering 
a certain evolution equation, as we are going to explain next; see \cite[Appendix C]{MK} 
for full details. Let 
\[ \Phi(u)={(Lu, u)}_Q={\|{\cal T}u\|}_Q^2-{({\cal K}{\cal T}u, u)}_Q \] 
be the functional in question. Strictly speaking, one considers $\Phi$ to be defined 
by the expression on the right-hand side, which makes sense for $u\in X^1_{{\rm odd}}$ only, 
but we will ignore this fact in what follows. 
For a given time interval $J=[0, a]$ or $J=[0, \infty[$ and a given continuous function 
$h: J\to X^1_{{\rm odd}}$ we introduce the family of operators 
\begin{eqnarray} 
   & & {\cal W}(t, s): g\mapsto {\cal W}(t, s)g,
   \quad {({\cal W}(t, s)g)}_k={\cal W}_k(t, s)g_k\,\,(k\in\Z), 
   \nonumber
   \\ & & {\cal W}_k(t, s)(I, \ell)
   =\exp\Big(-\int_s^t [k^2\omega_1^2(I, \ell)-\Phi(h(\tau))]\,d\tau\Big),
   \label{loezah2} 
\end{eqnarray} 
for $t, s\in J$, $t\ge s$; to emphasize the dependence on $h$, 
we will at times also write ${\cal W}(t, s; h)$. Note the evolution system property 
\[ {\cal W}(t, s)\circ {\cal W}(s, \tau)={\cal W}(t, \tau),\quad t, s, 
   \tau\in J,\,\,t\ge s\ge\tau. \] 
We will consider the evolution equation 
\begin{equation}\label{wosch} 
   g(t)={\cal W}(t, 0)\psi+\int_0^t {\cal W}(t, s)\,{\cal K}{\cal T}g(s)\,ds
\end{equation} 
for $t\ge 0$ and initial data $\psi$, where ${\cal W}(t, s)={\cal W}(t, s; g)$. 
For this evolution equation one can establish that if $\psi\in X^2_{{\rm odd}}$ 
is such that ${\|\psi\|}_Q=1$ and $\Phi(\psi)\le\lambda_\ast+\eps_\ast$ 
(for $\eps_\ast>0$ small enough), 
then there exists a global continuous solution $g: [0, \infty[\to X^1_{{\rm odd}}$ 
of (\ref{wosch}) that satisfies ${\|g(t)\|}_{X^0}=1$ for $t\in [0, \infty[$. 
This result does not rely on $\lambda_\ast<\delta_1^2$, 
the condition $\lambda_\ast\le\delta_1^2$ 
is enough. The point about (\ref{wosch}) is the following. Differentiating (\ref{loezah2}) 
for $h=g$ w.r.~to $t$, we get 
\[ \partial_t {\cal W}_k(t, s)(I, \ell)
   =-[k^2\omega_1^2(I, \ell)-\Phi(g(t))]\,{\cal W}_k(t, s)(I, \ell) \] 
and hence, at least formally, 
\[ \partial_t ({\cal W}(t, s)g)\cong ( \partial_t {\cal W}_k(t, s)g_k)
   =(-[k^2\omega_1^2-\Phi(g(t))]\,{\cal W}_k(t, s)g_k)
   \cong {\cal T}^2 {\cal W}(t, s)g+\Phi(g(t))\,{\cal W}(t, s)g. \] 
Applying this relation to (\ref{wosch}), it follows that  
\begin{eqnarray}\label{targ}  
   g'(t) & = & {\cal T}^2 {\cal W}(t, s)\psi+\Phi(g(t))\psi
   +\int_0^t [{\cal T}^2 {\cal W}(t, s){\cal K}{\cal T}g(s)
   +\Phi(g(t))\,{\cal W}(t, s){\cal K}{\cal T}g(s)]\,ds
   \nonumber
   \\ & & +\,{\cal K}{\cal T}g(t)
   \nonumber
   \\ & = & {\cal T}^2 g(t)+\Phi(g(t))\,g(t)+{\cal K}{\cal T}g(t)
   \nonumber
   \\ & = & -Lg(t)+\Phi(g(t))\,g(t). 
\end{eqnarray} 
This implies that the ${\|\cdot\|}_Q$-norm is preserved along the solution flow. 
Since $\Phi(u)={(Lu, u)}_Q$ for $u\in X^2_{{\rm odd}}$ and as the solution $g(t)$ 
is regular enough, we also deduce from (\ref{targ}) that 
\begin{eqnarray*} 
   \frac{d}{dt}\,\Phi(g(t)) & = & \frac{d}{dt}\,{(Lg(t), g(t))}_Q
   =2\,{(Lg(t), g'(t))}_Q
   \\ & = & 2\,{(Lg(t), -Lg(t)+\Phi(g(t))\,g(t))}_Q
   =-2\,({\|Lg(t)\|}^2_Q-\Phi(g(t))^2).
\end{eqnarray*} 
Now if ${\|g(0)\|}_Q=1$ initially, then 
\begin{eqnarray*} 
   {\|g'(t)\|}^2_Q
   & = & {\|-Lg(t)+\Phi(g(t))g(t)\|}^2_Q
   \\ & = & {\|Lg(t)\|}^2_Q-2\Phi(g(t))\,{(Lg(t), g(t))}_Q
   +\Phi(g(t))^2\,{\|g(t)\|}^2_Q
   \\ & = & {\|Lg(t)\|}^2_Q-\Phi(g(t))^2,
\end{eqnarray*} 
which in turn yields 
\[ \frac{d}{dt}\,\Phi(g(t))=-2\,{\|g'(t)\|}^2_Q\le 0. \] 
Therefore we see that $\Phi$ is a Lyapunov function for the evolution. 
Since ${\|g(t)\|}_Q=1$, we also have $\Phi(g(t))={(Lg(t), g(t))}_Q\ge\lambda_\ast$, 
and it is a natural question to ask, 
if we can construct a minimizer of $\Phi$ in the following way. 
Consider a sequence of initial data $(\psi_j)\subset X^2_{{\rm odd}}$ 
such that $\Phi(\psi_j)\le\lambda_\ast+1/j$ and let $g_j$ denote the corresponding solution 
to (\ref{wosch}) so that $g_j(0)=\psi_j$. Then $\lambda_\ast\le\Phi(g_j(t))
\le\Phi(\psi_j)\le\lambda_\ast+1/j$ 
for all $t\in [0, \infty[$ and $j\in\N$. Hence the key point is to find a sequence 
of times $(t_j)$ 
with the properties that $t_j\to\infty$ and $\{g_j(t_j): j\in\N\}\subset X^0$ 
is relatively compact. It can be shown that this goal can be accomplished, 
if the condition $\lambda_\ast<\delta_1^2$ is imposed; the limiting function $u_\ast$ 
will then be the desired eigenfunction of $L$ for the eigenvalue $\lambda_\ast$. 
 
%%%%%%%%%%%%%%%%%%%%%%%%%%%%%%%%%%%%%%%%%%%%%%%%%%%%%%%%%%%%%%%%%%%%%%%%%%%%%%% 
 
\section{Open questions and further topics} 
\setcounter{equation}{0} 
 
\begin{itemize}
\item Do some numerics. 
\item Can it happen, for some static solution, that $\lambda_\ast=\delta_1^2$? 
\item Determine where $\omega_1$ attains its minimum on $D$. 
Is it the same point for all ``reasonable'' static solutions $Q$? 
\item Determine the limit $\mu_\ast$ from (\ref{iastdef}) 
in terms of $Q$. 
\item When it comes to relativistic galactic dynamics, the appropriate model 
is the Einstein-Vlasov system 
\cite{Andreasson_review}. In the present lectures we have not been dealing with this 
more general system, but of course it will be tempting to determine 
which results 
could be transferred to Einstein-Vlasov; see
\cite{IpsTho1,Ips2,Ips3,Facker4,Facker5,HaRe1,HaRe2} for work 
in this context that is related to the Antonov bound. 
\end{itemize} 

%%%%%%%%%%%%%%%%%%%%%%%%%%%%%%%%%%%%%%%%%%%%%%%%%%%%%%%%%%%%%%%%%%%%%%%%%%%%%%%


\begin{thebibliography}{99}

\bibitem{Andreasson_review}
 {\sc Andr\'{e}asson H.:} The Einstein-Vlasov system/kinetic theory, 
 {\em Living Rev.~Relativ.}{\bf\,\,5}, 33 pp., 2002-7 (2002)
\bibitem{anton1} 
 {\sc Antonov V.A.:} Remarks on the problem of stability in stellar dynamics, 
 (in Russian), {\em Astronom.~\v{Z}.}{\bf\,\,37}, 918-926 (1960); translated in 
 {\em Soviet Astronom.~AJ}{\bf\,\,4}, 859-867 (1960) 
\bibitem{anton2} 
 {\sc Antonov V.A.:} Solution of the problem of stability of a stellar system 
 with the Emden density law and spherical velocity distribution, (in Russian), 
 {\em J.~Leningr.~Univ.~Ser.~Mekh.~Astron.}{\bf\,\,7}, 135-146 (1962) 
\bibitem{BT}
 {\sc Binney J.\,\& Tremaine S.:}{\em\,\,Galactic Dynamics},
 2nd edition, Princeton University Press, Princeton 2008
\bibitem{bir} 
 {\sc Birman M.\v{S}.:} On the spectrum of singular boundary-value problems, 
 (in Russian), {\em Mat.~Sb.~(N.S.)}{\bf\,\,55 (97)}, 125-174 (1961); 
 translated in {\em Amer.~Math.~Soc.~Transl.}{\bf\,\,53}, 23-80 (1966) 
\bibitem{DBF} 
 {\sc Doremus J.P., Baumann G.\,\& Feix M.R.:} Stability of a self gravitating system 
 with phase space density function of energy and angular momentum, 
 {\em Astronom. and Astrophys.}{\bf\,\,29}, 401-407 (1973) 
\bibitem{DL} 
 {\sc Dyson F.J.\,\& Lenard A.:} Stability of matter. I and II, 
 {\em J.~Math.~Phys.}{\bf\,\,8}, 423-434 (1967); ibid.{\bf\,\,9}, 698-711 (1968)
\bibitem{Facker4} 
 {\sc Fackerell E.D.:} Relativistic, spherically symmetric star clusters. IV. 
 A sufficient condition for instability of isotropic clusters against radial perturbations, 
 {\em Astrophys.~J.}{\bf\,\,160}, 859-874 (1970) 
\bibitem{Facker5} 
 {\sc Fackerell E.D.:} Relativistic, spherically symmetric star clusters. V. 
 A relativistic version of Plummer's model, 
 {\em Astrophys.~J.}{\bf\,\,165}, 489-493 (1971) 
\bibitem{GCDB} 
 {\sc Gillon D., Cantus M., Doremus J.P.\,\& Baumann G.:} 
 Stability of self-gravitating spherical systems 
 in which phase space density is a function of energy and angular momentum, 
 for spherical perturbations, 
 {\em Astronom. and Astrophys.}{\bf\,\,50}, 467-470 (1976) 
\bibitem{Glassey_book} 
 {\sc Glassey R.T.:}{\em\,\,The Cauchy Problem in Kinetic Theory}, SIAM, Philadelphia 1996
\bibitem{GuoAnt} 
 {\sc Guo Y.:} On the generalized Antonov stability criterion, 
 in {\em Nonlinear Wave Equations (Providence, RI, 1998)}, 
 Contemp.~Math.~{\bf 263}, American Mathematical Society, Providence, 85-107 (2000)
\bibitem{GRking}
 {\sc Guo Y.\,\& Rein G.:} A non-variational approach to nonlinear stability 
 in stellar dynamics 
 applied to the King model, {\em Comm.~Math.~Phys.}{\bf\,\,271}, 489-509 (2007)  
\bibitem{HaRe1} 
 {\sc Had\v{z}i\'{c} M.\,\& Rein G.:} Stability for the spherically symmetric 
 Einstein-Vlasov system-a coercivity estimate, 
 {\em Math.~Proc.~Cambridge Philos.~Soc.}{\bf\,\,155}, 529-556 (2013) 
\bibitem{HaRe2} 
 {\sc Had\v{z}i\'{c} M.\,\& Rein G.:} On the small redshift limit of steady states 
 of the spherically symmetric Einstein-Vlasov system and their stability, 
 {\em Math.~Proc.~Cambridge Philos.~Soc.}{\bf\,\,159}, 529-546 (2015)
\bibitem{BAH} 
 {\sc Had\v{z}i\'{c} M., Rein G.\,\& Straub C.:} On the existence of linearly 
 oscillating galaxies, {\em Arch.~Rational Mech.~Anal.}{\bf\,\,243}, 611-696 (2022)  
\bibitem{Henon}
 {\sc H\'{e}non M.:} Vlasov equation?, 
 {\em Astron.~Astrophys.}{\bf\,\,114}, 211-212 (1982)
\bibitem{HiSi}
 {\sc Hislop P.D.\,\& Sigal I.M.:}{\em\,\,Introduction to Spectral Theory.
 With Applications to Schr\"odinger Operators}, Springer, Berlin-New York 1996
\bibitem{IpsTho1} 
 {\sc Ipser J.R.\,\& Thorne K.S.:} Relativistic, spherically symmetric star clusters. I. 
 Stability theory for radial perturbations, 
 {\em Astrophys.~J.}{\bf\,\,154}, 251-270 (1968)  
\bibitem{Ips2} 
 {\sc Ipser J.R.:} Relativistic, spherically symmetric star clusters. II. 
 Sufficient conditions for stability against radial perturbations, 
 {\em Astrophys.~J.}{\bf\,\,156}, 509-527 (1969) 
\bibitem{Ips3} 
 {\sc Ipser J.R.:} Relativistic, spherically symmetric star clusters. III. 
 Stability of compact isotropic models, 
 {\em Astrophys.~J.}{\bf\,\,158}, 17-43 (1969) 
\bibitem{Jeans_paper} 
 {\sc Jeans J.H.:} On the theory of star-streaming and the structure of the universe, 
 {\em Monthly Notices Roy.~Astronom.~Soc.}{\bf\,\,76}, 70-84 (1915) 
\bibitem{Kand} 
 {\sc Kandrup H.E.:} A stability criterion for any collisionless stellar equilibrium 
 and some concrete 
 applications thereof, {\em Astrophys.~J.}{\bf\,\,370}, 312-317 (1991)  
\bibitem{KS}  
 {\sc Kandrup H.E.\,\& Sygnet J.F.:} A simple proof of dynamical stability 
 for a class of spherical clusters, 
 {\em Astrophys.~J.}{\bf\,\,298}, 27-33 (1985) 
\bibitem{MK}
 {\sc Kunze M.:}{\em\,\,A Birman-Schwinger Principle in Galactic Dynamics}, 
 Birkh\"auser/Springer, Cham 2021 
\bibitem{LMR11}
 {\sc Lemou M., M\'{e}hats F.\,\& Rapha\"{e}l P.:} A new variational approach 
 to the stability 
 of gravitational systems, {\em Comm.~Math.~Phys.}{\bf\,\,302}, 161-224 (2011) 
\bibitem{LMR12}
 {\sc Lemou M., M\'{e}hats F.\,\& Rapha\"{e}l P.:} Orbital stability 
 of spherical galactic models, 
 {\em Invent.~Math.}{\bf\,\,187}, 145-194 (2012)  
\bibitem{LS}
 {\sc Lieb E.H.\,\& Seiringer R.:}{\em\,\,The Stability of Matter in Quantum Mechanics}, 
 Cambridge University Press, Cambridge 2010
\bibitem{LT} 
 {\sc Lieb E.H.\,\& Thirring W.:} Bound for the kinetic energy of fermions which 
 proves the stability of matter, {\em Phys.~Rev.~Lett.}{\bf\,\,35}, 687-689 (1975); 
 Errata ibid., 1116 (1975)
\bibitem{LP} 
 {\sc Lions P.-L.\,\& Perthame B:} Propagation of moments and regularity 
 for the 3-dimensional Vlasov-Poisson system, 
 {\em Invent.~Math.}{\bf\,\,105} 415-430 (1991) 
\bibitem{MP} 
 {\sc Mar\'{e}chal L.\,\& Perez J.:} Radial orbit instability 
 as a dissipation-induced phenomenon, 
 {\em Monthly Notices Roy.~Astronom.~Soc.}{\bf\,\,405}, 2785-2790 (2010) 
\bibitem{mathur}
 {\sc Mathur S.D.:} Existence of oscillation modes in collisionless gravitating systems,
 {\em Monthly Notices Roy.~Astronom.~Soc.}{\bf\,\,243}, 529-536 (1990)
\bibitem{morrison} 
 {\sc Morrison P.J.:} Hamiltonian description of the ideal fluid, 
 {\em Rev.~Modern Phys.}{\bf\,\,70}, 467-521 (1998) 
\bibitem{Mouhot_rev} 
 {\sc Mouhot C.:} Stabilit\'{e} orbitale pour le syst\`{e}me de Vlasov-Poisson gravitationnel 
 (d'apr\`{e}s Lemou-M\'{e}hats-Rapha\"{e}l, Guo, Lin, Rein et al.), Ast\'{e}risque {\bf 352}, 
 35-82 (2013) 
\bibitem{Perez} 
 {\sc Perez J.\,\& Aly J.-J..:} Stability of spherical stellar systems-I. Analytical results, 
 {\em Monthly Notices Roy.~Astronom.~Soc.}{\bf\,\,280}, 689-699 (1996)
\bibitem{pfaff}
 {\sc Pfaffelmoser K.:} Global classical solutions of the Vlasov-Poisson system 
 in three dimensions 
 for general initial data, {\em J.~Differential Equations}{\bf\,\,95}, 281-303 (1992) 
\bibitem{RSIV}
 {\sc Reed M.\,\& Simon B.:}{\em\,\,Methods of Modern Mathematical
 Physics IV: Analysis of Operators}, Academic Press, New York 1978
\bibitem{Reinrev}
 {\sc Rein G.:} Collisionless kinetic equations from astrophysics--the Vlasov-Poisson system, 
 in {\em Handbook of Differential Equations: Evolutionary Equations. Vol. III}, 
 Elsevier/North-Holland, Amsterdam 2007, pp.~383-476
\bibitem{schaeff} 
 {\sc Schaeffer J.:} Global existence of smooth solutions to the Vlasov-Poisson system 
 in three dimensions, {\em Comm.~Partial Differential Equations}{\bf\,\,16}, 1313-1335 (1991)
\bibitem{schwi} 
 {\sc Schwinger J.:} On the bound states of a given potential, 
 {\em  Proc.~Nat.~Acad.~Sci.~U.S.A.}{\bf\,\,47}, 122-129 (1961)
\bibitem{Sim_quadr}
 {\sc Simon B.:}{\em\,\,Quantum Mechanics for Hamiltonians Defined as Quadratic Forms}, 
 Princeton University Press, Princeton 1971
\bibitem{FunctInt}
 {\sc Simon B.:}{\em\,\,Functional Integration and Quantum Physics}, 
 Academic Press, New York-London 1979
\bibitem{Simon4}
 {\sc Simon B.:}{\em\,\,A Comprehensive Course in Analysis, Part 4: Operator Theory}, 
 American Mathematical Society, Providence 2015 
\bibitem{SdLRP}  
 {\sc Sygnet J.F., des Forets G., Lachieze-Rey M.\,\& Pellat R.:} 
 Stability of gravitational systems 
 and gravothermal catastrophe in astrophysics, {\em Astrophys.~J.}{\bf\,\,276}, 
 737-745 (1984) 
\bibitem{thirr}
 {\sc Thirring W.:}{\em\,\,Lehrbuch der Mathematischen Physik,
 Band 1: Klas\-si\-sche Dynamische Sy\-steme}, 2nd edition, Springer,
 Berlin-New York 1988
\bibitem{Vlasov_paper} 
 {\sc Vlasov A.A.:} The vibrational properties of the electron gas, 
 {\em Zh.~Eksp.~Teor.~Fiz.}{\bf\,\,8}, 291 (1938) and {\em Usp.~Fiz.~Nauk}{\bf\,\,93}, 
 444 (1967); 
 see {\em Sov.~Phys.~Usp.}{\bf\,\,10}, 721 (1968)
\bibitem{zehnder}
 {\sc Zehnder E.:}{\em\,\,Lectures on Dynamical Systems}, 
 European Mathematical Society, Z\"urich 2010 

\end{thebibliography}
\end{document}